\documentclass[twocolumn,english,superscriptaddress,showpacs,prl]{revtex4}
\usepackage[latin9]{inputenc}
\usepackage{color}
\usepackage{babel}
\usepackage{float}
\usepackage{bm}
\usepackage{amsmath}
\usepackage{amssymb}
\usepackage{graphicx}
\usepackage{esint}
\usepackage[unicode=true,pdfusetitle,
 bookmarks=true,bookmarksnumbered=true,bookmarksopen=false,
 breaklinks=false,pdfborder={0 0 1},backref=false,colorlinks=true]
 {hyperref}
\hypersetup{
 citecolor=blue}

\makeatletter

\@ifundefined{textcolor}{}
{%
 \definecolor{BLACK}{gray}{0}
 \definecolor{WHITE}{gray}{1}
 \definecolor{RED}{rgb}{1,0,0}
 \definecolor{GREEN}{rgb}{0,1,0}
 \definecolor{BLUE}{rgb}{0,0,1}
 \definecolor{CYAN}{cmyk}{1,0,0,0}
 \definecolor{MAGENTA}{cmyk}{0,1,0,0}
 \definecolor{YELLOW}{cmyk}{0,0,1,0}
}

\usepackage{times}

\makeatother

\begin{document}

\title{Kaleidoscope of quantum phases in a long-range interacting spin-1
chain}

\author{Z.\,-X. Gong}
\email{gzx@umd.edu}

\affiliation{Joint Quantum Institute, NIST/University of Maryland, College Park,
Maryland 20742, USA}

\affiliation{Joint Center for Quantum Information and Computer Science, NIST/University
of Maryland, College Park, Maryland 20742, USA}

\author{M.\,F. Maghrebi}

\affiliation{Joint Quantum Institute, NIST/University of Maryland, College Park,
Maryland 20742, USA}

\affiliation{Joint Center for Quantum Information and Computer Science, NIST/University
of Maryland, College Park, Maryland 20742, USA}

\author{A. Hu}

\affiliation{Joint Quantum Institute, NIST/University of Maryland, College Park,
Maryland 20742, USA}

\affiliation{Department of Physics, American University, Washington, DC 20016,
USA}

\author{M. Foss-Feig}

\affiliation{Joint Quantum Institute, NIST/University of Maryland, College Park,
Maryland 20742, USA}

\affiliation{Joint Center for Quantum Information and Computer Science, NIST/University
of Maryland, College Park, Maryland 20742, USA}

\author{P. Richerme}

\affiliation{Joint Quantum Institute, NIST/University of Maryland, College Park,
Maryland 20742, USA}

\affiliation{Department of Physics, Indiana University, Bloomington, Indiana,
47405, USA}

\author{C. Monroe}

\affiliation{Joint Quantum Institute, NIST/University of Maryland, College Park,
Maryland 20742, USA}

\affiliation{Joint Center for Quantum Information and Computer Science, NIST/University
of Maryland, College Park, Maryland 20742, USA}

\author{A.\,V. Gorshkov}

\affiliation{Joint Quantum Institute, NIST/University of Maryland, College Park,
Maryland 20742, USA}

\affiliation{Joint Center for Quantum Information and Computer Science, NIST/University
of Maryland, College Park, Maryland 20742, USA}
\begin{abstract}
Motivated by recent trapped-ion quantum simulation experiments, we
carry out a comprehensive study of the phase diagram of a spin-1 chain
with XXZ-type interactions that decay as $1/r^{\alpha}$, using a
combination of finite and infinite-size DMRG calculations, spin-wave
analysis, and field theory. In the absence of long-range interactions,
varying the spin-coupling anisotropy leads to four distinct phases:
a ferromagnetic Ising phase, a disordered XY phase, a topological
Haldane phase, and an antiferromagnetic Ising phase. If long-range
interactions are antiferromagnetic and thus frustrated, we find primarily
a quantitative change of the phase boundaries. On the other hand,
ferromagnetic (non-frustrated) long-range interactions qualitatively
impact the entire phase diagram. Importantly, for $\alpha\lesssim3$,
long-range interactions destroy the Haldane phase, break the conformal
symmetry of the XY phase, give rise to a new phase that spontaneously
breaks a $U(1)$ continuous symmetry, and introduce an exotic tricritical
point with no direct parallel in short-range interacting spin chains.
We show that the main signatures of all five phases found could be
observed experimentally in the near future. 
\end{abstract}


\maketitle
The study of quantum phase transitions in low-dimensional spin systems
has been a major theme in condensed matter physics for many years
\cite{sachdev_quantum_2011}. A well-known implication of Mermin and
Wagner's famous results \cite{mermin_absence_1966} on finite temperature
quantum systems is that, for a large class of one-dimensional quantum
spin systems, long-range order is forbidden even at zero temperature.
This absence of classical order promotes quantum fluctuations to a
central role, and they often determine the qualitative properties
of the quantum ground state. An important example, first conjectured
by Haldane \cite{haldane_continuum_1983,haldane_nonlinear_1983},
is that a spin-1 antiferromagnetic Heisenberg chain possesses a disordered
phase with an energy gap to bulk excitations, later identified as
a symmetry protected topological phase \cite{pollmann_symmetry_2012,chen_symmetry_2013}.
Its spin-1/2 counterpart, despite possessing the same classical limit,
has a disordered ground state with gapless excitations, and is described
by a conformal field theory (CFT) \cite{difrancesco_conformal_1997}.

Experimentally, such quantum phase transitions have been explored
in quasi-1D materials, and more recently using artificial materials
designed through the careful control of atomic, molecular, and optical
(AMO) systems \cite{bloch_manybody_2008,kim_quantum_2010,bloch_quantum_2012,britton_engineered_2012}.
These AMO systems are usually well-isolated from the environment,
of system parameters, and make possible both measurement and control
at the individual lattice-site level. A distinctive feature of AMO
systems is that interactions between particles are often long-ranged,
decaying as a power-law with distance ($1/r^{\alpha}$). The exponent
$\alpha$ varies widely amongst different AMO systems, ranging from
$\alpha=6$ for van de Waals interactions in Rydberg atoms, to $\alpha=3$
for polar molecules and magnetic atoms, to $\alpha=0$ for atoms coupled
to cavities \cite{yan_observation_2013,peter_anomalous_2012,douglas_quantum_2015,britton_engineered_2012,jurcevic_quasiparticle_2014,hazzard_manybody_2014,depaz_nonequilibrium_2013,richerme_nonlocal_2014,schauss_observation_2012}.
The effect of long-range interactions can be tuned by either changing
the dimensionality of the system, e.g. for neutral atoms or molecules
in optical lattices, or by directly (and often continuously) altering
the value of $\alpha$, e.g. in trapped ions or cold atoms coupled
to photonic crystals \cite{douglas_quantum_2015}. The availability
of tunable long-range interactions creates an entirely new degree
of freedom\textemdash absent in typical condensed-matter systems\textemdash for
inducing quantum phase transitions, and can potentially lead to novel
quantum phases \cite{manmana_topological_2013,yao_topological_2012,yao_realizing_2013}.

While long-range interacting classical models have been studied in
considerable detail for some time \cite{fisher_critical_1972,cannas_longrange_1996,luijten_classical_1997,katzgraber_probing_2005,campa_statistical_2009},
there is a relative lack of in-depth studies of quantum phase transitions
in long-range interacting systems, despite the emerging experimental
prospects for studying both their equilibrium and non-equilibrium
properties \cite{koffel_entanglement_2012,bachelard_universal_2013,eisert_breakdown_2013,gong_persistence_2014,jurcevic_quasiparticle_2014,vodola_kitaev_2014,richerme_nonlocal_2014,hazzard_manybody_2014,foss-feig_nearly_2015,vodola_longrange_2015}.
One reason is that many analytically solvable lattice models become
intractable when interactions are no longer short-ranged, a well-known
example being the spin-1/2 XXZ model. In addition, to properly incorporate
long-range interactions in low-energy effective theories, existing
field theoretic treatments need to be modified and usually become
much more complicated \cite{maghrebi_causality_2015,rajabpour_quantum_2015}.
Though numerically exact techniques for quantum systems have been
adapted to treat long-range interactions, significant challenges remain
in the numerical calculation of phase diagrams. In particular, power-law
decaying interactions generally lead to a divergent correlation length
\cite{hastings_spectral_2006,gong_persistence_2014}, and a much larger
system size or a much higher precision is typically required to faithfully
describe the properties of the system in the long-wavelength limit.
Several authors have performed analytical studies of non-interacting
bosonic and fermionic systems with long-range hopping and pairing
\cite{nezhadhaghighi_quantum_2013,nezhadhaghighi_entanglement_2012,vodola_kitaev_2014,vodola_longrange_2015},
but there have been relatively few numerical studies of non-integrable
systems, and those that exist have primarily focused on spin-1/2 chains
\cite{laflorencie_critical_2005,sandvik_ground_2010,koffel_entanglement_2012,manmana_topological_2013,blanchard_influence_2013}.

In this manuscript, we carry out a detailed study of a spin-1 chain
with tunable XXZ interactions that decay monotonically as $1/r^{\alpha}$,
for all $\alpha>0$. Our study is largely motivated by imminent trapped-ion
based experiments that can simulate this model with widely tunable
index $\alpha$ \cite{cohen_proposal_2014,cohen_simulating_2015,senko_realization_2015}.
In the absence of long-range interactions, the choice of spin-1 over
spin-1/2 allows us to have four distinct quantum phases by varying
the anisotropy of the interactions: a ferromagnetic (FM) phase and
an antiferromagnetic (AFM) Ising phase that are both gapped and long-range
ordered, a disordered gapless phase (the XY phase), and a gapped and
topologically ordered phase (the Haldane phase). By using a combination
of density matrix renormalization group (DMRG) calculations, spin
wave analysis, and field theory, we obtain the phase diagram for arbitrary
anisotropy and all $\alpha>0$, with both ferromagnetic and antiferromagnetic
interactions. Our key observation is that, when interactions in all
spatial directions are antiferromagnetic, long-range interactions
are frustrated, leading to primarily quantitative changes to the phase
boundaries compared to the short-range interacting chain. Interestingly,
we find that the topological Haldane phase is robust under long-range
interactions with any $\alpha>0$ \cite{gong_topological_2015}. However,
when the interactions in the $x-y$ plane become ferromagnetic, we
find a number of significant modifications to the phase diagram: (1)
The Haldane phase is destroyed at a finite $\alpha$ due to a closing
of the bulk excitation gap; (2) The gapless XY phase, described by
a CFT with central charge $c=1$, disappears when $\alpha\lesssim3$
due to a breakdown of conformal symmetry \cite{vodola_kitaev_2014,vodola_longrange_2015};
(3) The disappearance of the XY phase heralds the emergence of a new
phase at $\alpha\lesssim3$ (continuous-symmetry breaking, or CSB)
in which the spins order in the $xy$ plane, spontaneously breaking
a $U(1)$ symmetry and possessing gapless excitations (Nambu-Goldstone
modes); (4) Novel tricritical points, with no direct analogue in short-range
interacting 1D models, appear at the intersection of the Haldane,
CSB, and XY/AFM phases.

The manuscript is organized as follows. In Sec.\ I, we introduce
the model Hamiltonian and present complete phase diagrams for the
ferromagnetic and antiferromagnetic cases. In Sec.\ II, we study
the boundary of the FM phase, where a spin-wave approximation is found
to be asymptotically exact in the large-system limit. In Sec.\ III,
we determine both the XY-to-Haldane and Haldane-to-AFM transition
lines accurately using DMRG calculations, and use field theory arguments
to explain the effect of long-range interactions on the boundary of
the Haldane phase. In Sec.\ IV, we introduce the new CSB phase and
explain its emergence using spin-wave theory. The boundary between
the CSB and XY phases is determined by a numerical calculation of
central charge. In Sec.\ V, we show that all five phases possess
distinct signatures that could be observed in near-future trapped
ion quantum simulations with chains of 16 spins. Finally, we conclude
the work in Sec.\ VI and comment on a number of open questions.

\section{Model Hamiltonian and phase diagrams}

We consider the following spin-1 Hamiltonian with long-range XXZ interactions
in a 1D open-boundary chain: 
\begin{eqnarray}
H & = & \sum_{i>j}\frac{1}{(i-j)^{\alpha}}[J_{xy}(S_{i}^{x}S_{j}^{x}+S_{i}^{y}S_{j}^{y})+J_{z}S_{i}^{z}S_{j}^{z}].\label{eq:H}
\end{eqnarray}
Here $J_{z}\in(-\infty,\infty)$ and $\alpha\in(0,\infty)$ are allowed
to vary continuously, and we consider both the $J_{xy}=1$ (anti-ferromagnetic)
and $J_{xy}=-1$ (ferromagnetic) cases. We note that, for $0<\alpha<1$,
Eq.\,\eqref{eq:H} does not have a well-defined thermodynamic limit
when $J_{xy}$ and/or $J_{z}$ is ferromagnetic, since the ground-state
energy-density diverges. To make the ground-state energy extensive,
we may impose an energy renormalization factor $N^{\alpha-1}$, first
introduced by Kac \cite{kac_van_1963}, when taking the thermodynamic
or continuum limit (here $N$ is the chain length). For finite-size
numerical calculations, we do not need to implement the Kac renormalization
for $0<\alpha<1$ since ground-state properties are unaffected by
energy renormalization \cite{InfKac}.

\begin{figure}
\includegraphics[width=0.4\textwidth]{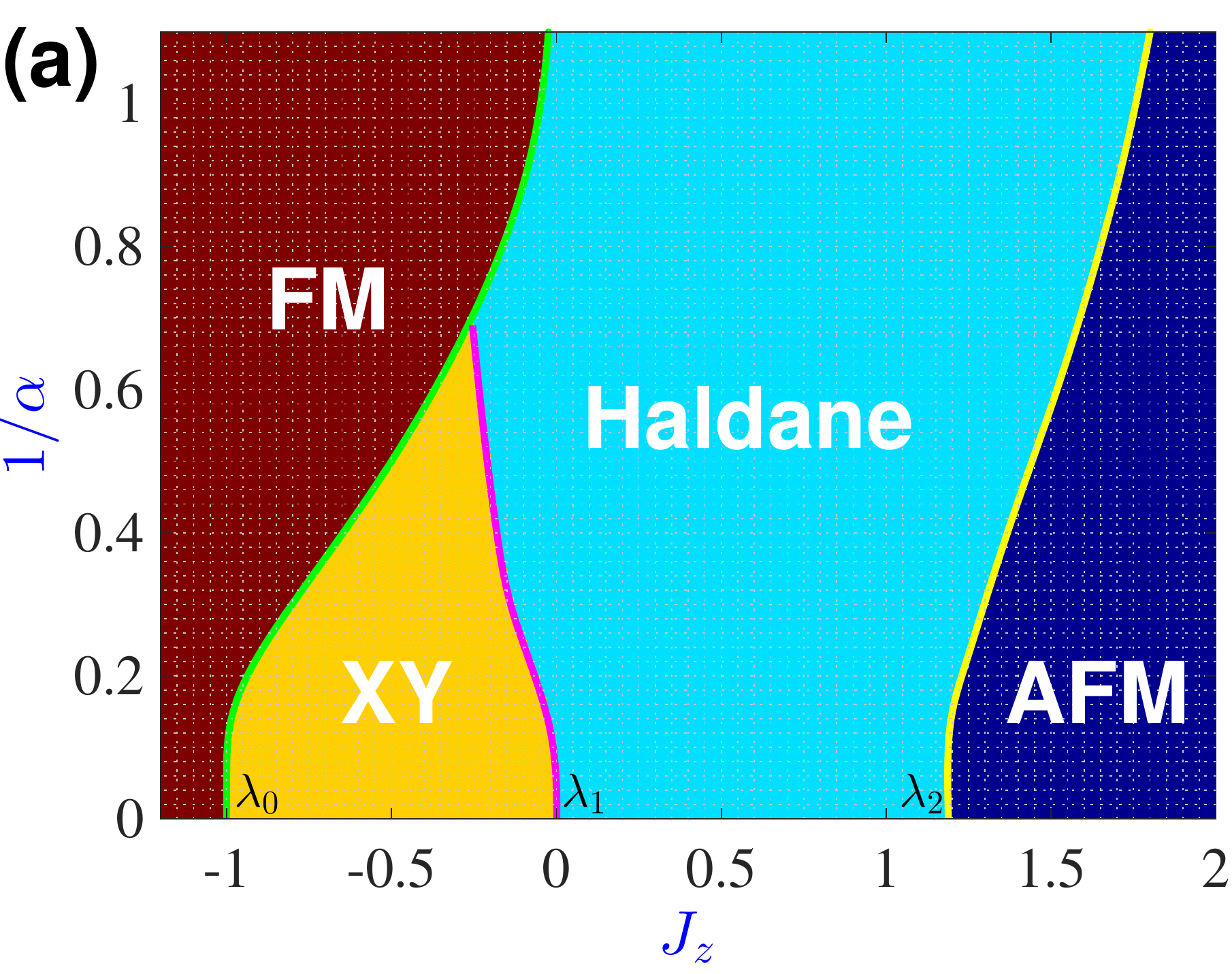}

\includegraphics[width=0.4\textwidth]{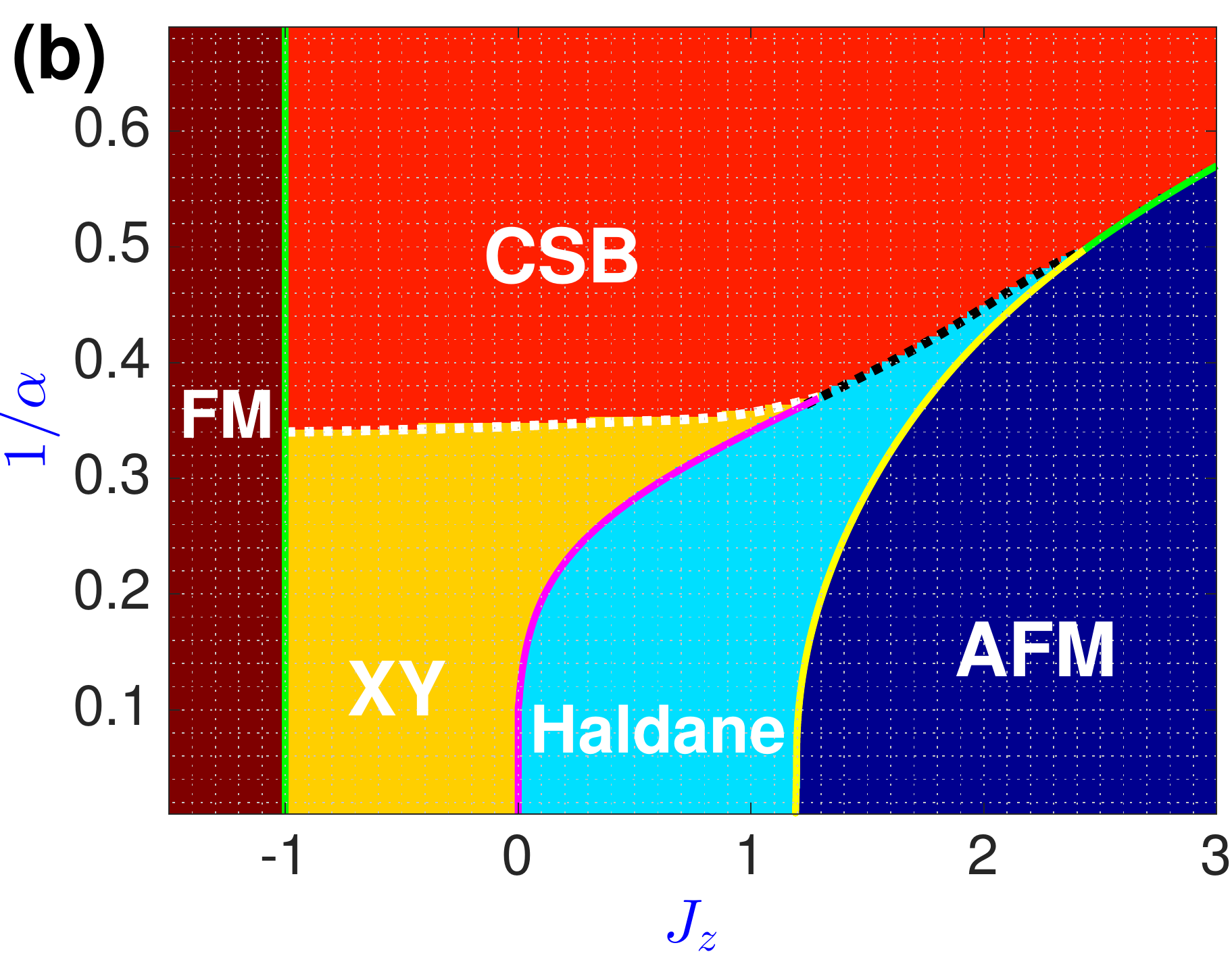}

\caption{\label{fig:PhaseDiagram}Proposed phase diagram for (a) $J_{xy}=1$
and (b) $J_{xy}=-1$. Five different phases are identified: a ferromagnetic
(FM) Ising phase, an antiferromagnetic (AFM) Ising phase, a disordered
XY phase, a topological Haldane phase, and a continuous symmetry breaking
(CSB) phase. At $\alpha=\infty$, the transition points are denoted
by $J_{z}=\lambda_{0,1,2}$ in (a). The FM-to-XY, FM-to-CSB, and CSB-to-AFM
transitions are first order (green line); the XY-to-Haldane transition
is BKT type with central charge $c=1$ (purple line); the Haldane-to-AFM
transition is second order with $c=0.5$ (yellow line); the CSB-to-XY
transition (white dashed line) has $c=1$, but is a BKT-like transition
corresponding to a universality class different from the XY-to-Haldane
transition \cite{maghrebi_continuous_2015}; the CSB-to-Haldane transition
(black dashed lines) appears to be an exotic continuous phase transition
not described by a 1+1D CFT. The location of solid transition lines
are expected to be accurate in the thermodynamic limit, while the
location of dashed transition lines may be inaccurate due to finite-size
effects in our numerics.}
\end{figure}

Figure\,\ref{fig:PhaseDiagram} shows our full phase diagram for
both $J_{xy}=1$ and $J_{xy}=-1$, with actual phase boundaries plotted
using the results of calculations discussed in the following sections.
The nearest-neighbor interaction limit is achieved at $\alpha\rightarrow\infty$
($1/\alpha=0$). In this limit, the Hamiltonian in Eq.\,\eqref{eq:H}
with $J_{xy}=-1$ is equivalent to the one with $J_{xy}=1$, as can
be seen by performing a local unitary transformation that flips every
other spin in the $x-y$ plane while preserving the spin commutation
relations: $S_{i}^{x,y}\rightarrow(-1)^{i}S_{i}^{x,y}$. The different
ground-state phases of this short-range Hamiltonian have been well-studied
\cite{kennedy_hidden_1992,kitazawa_phase_1996,tsukano_spin1_1998}.
Notably, Haldane first conjectured \cite{haldane_continuum_1983,haldane_nonlinear_1983}
that for $\lambda_{1}<J_{z}<\lambda_{2}$, a disordered gapped phase
(the Haldane phase) will emerge. At $J_{z}=\lambda_{2}$, the ground
state undergoes a second-order phase transition from the Haldane phase
to an AFM phase, which belongs to the same universality class as the
2D Ising model. The value $\lambda_{2}\approx1.186$ has been found
by various numerical techniques including Monte-Carlo \cite{nomura_spincorrelation_1989},
exact diagonalization \cite{sakai_finitesize_1990}, and DMRG \cite{su_nonlocal_2012,liu_entanglement_2014,ejima_comparative_2015}.
At $J_{z}=\lambda_{1}$, a Berezinskii-Kosterlitz-Thouless (BKT) transition
intervenes between the Haldane phase and a gapless disordered XY phase
at $J_{z}<\lambda_{1}$. The value of $\lambda_{1}$ is theoretically
predicted to be exactly zero after mapping Eq.\,\eqref{eq:H} (for
$\alpha=\infty$) to a field theory model using bosonization \cite{schulz_phase_1986}.
This prediction is supported by conformal field theory arguments \cite{alcaraz_critical_1992}
and a level spectroscopy method based on a renormalization group analysis
and the SU$(2)/Z_{2}$ symmetry of the BKT transition \cite{nomura_correlation_1995,kitazawa_phase_1996,nomura_su2z2_1998,kitazawa_su2_2003}.
Numerically, $\lambda_{1}\approx0$ has been verified via finite-size
scaling \cite{botet_groundstate_1983,sakai_finitesize_1990,ueda_finitesize_2008}
and DMRG \cite{su_nonlocal_2012}. Finally, at $J_{z}=\lambda_{0}=-1$,
a first-order phase transition from the XY phase to a ferromagnetic
Ising phase takes place \cite{kitazawa_phase_1996,chen_groundstate_2003,liu_entanglement_2014}.

We now introduce our results for the long-range interacting case ($1/\alpha>0$).
For $J_{xy}=1$ and $J_{z}>0$, long-range interactions are frustrated
and the Haldane-to-AFM phase transition point $\lambda_{2}(\alpha)$
increases moderately as $\alpha$ decreases, without changing the
universality class of the transition. For sufficiently small $J_{z}<0$
, the ferromagnetic long-range interactions along the $z$ direction
eventually favor a ferromagnetic ground state, inducing a first-order
transition at $\lambda_{0}(\alpha)$. The magnitude of the critical
coupling, $|\lambda_{0}(\alpha)|$, decreases monotonically from $1$
(at $\alpha=\infty$) to $0$ (for all $\alpha\le1$) in the thermodynamic
limit. The XY-to-Haldane phase boundary $\lambda_{1}(\alpha)$ becomes
negative for finite $\alpha$, similar to the XXZ spin-1 chain with
next-nearest-neighbor interactions \cite{nomura_phase_1993}, eventually
terminating in a tricritical point at the intersection of FM, Haldane,
and XY phases. The entire XY phase (including the XY-to-Haldane phase
boundary) has conformal symmetry with $c=1$, and the XY-to-Haldane
phase boundary remains a BKT transition until it terminates at the
tricritical point.

For $J_{xy}=-1$, where long-range interactions in the $x-y$ plane
are not frustrated, the phase diagram {[}Fig.\,\ref{fig:PhaseDiagram}(b){]}
shows a number of important qualitative differences from the nearest-neighbor
phase diagram as $\alpha$ is decreased. First, the XY-to-Haldane
phase boundary bends significantly toward positive $J_{z}$, and we
find the Haldane phase to terminate at $\alpha\approx3$ for $J_{z}=1$
. Second, we expect the XY phase to disappear for $\alpha\lesssim3$
due to the breakdown of conformal symmetry \cite{vodola_kitaev_2014,vodola_longrange_2015}.
Third, for $\alpha\lesssim3$ a new CSB phase emerges\textemdash this
is not in violation of the Mermin-Wagner theorem, as it no longer
applies for this range of interactions \cite{mermin_absence_1966,bruno_absence_2001,laflorencie_critical_2005,Lobos_restoring_2013,Tezuka_destruction_2014,maghrebi_continuous_2015}.
The CSB-to-AFM phase transition is expected to be first-order, since
at large $J_{z}$ and small $\alpha$, quantum fluctuations play negligible
roles for both the Néel-ordered state and the ordered CSB state. This
behavior is similar to the transition between the AFM phase and the
large-$D$ phase (where a large positive anisotropy term $D\sum_{i}(S_{i}^{z})^{2}$
causes all spins to stay in the $|S_{i}^{z}=0\rangle$ state) reported
in Refs.\,\cite{chen_groundstate_2003,su_nonlocal_2012,liu_entanglement_2014}.
The Haldane phase has a $c=1$ critical phase boundary with the XY
phase, a $c=0.5$ phase boundary with the AFM phase \cite{ejima_comparative_2015},
and a possibly exotic phase boundary with the CSB phase, a boundary
that is not described by a 1+1D CFT.

\section{FM phase and its boundary}

Because the ferromagnetic state with all spins polarized along $\pm z$
(or an arbitrary superposition of these two states) is an exact eigenstate
of the Hamiltonian for any value of $\alpha$ and $J_{z}$, we expect
a first-order quantum phase transition at the boundary of the FM phase.
The FM state has an energy $E_{{\rm FM}}=J_{z}\sum_{i>j}(i-j)^{-\alpha}$,
and the phase transition out of this state, defining the critical
line $J_{z}=\lambda_{0}(\alpha)$, occurs when some other eigenstate
with no ferromagnetic order appears with a lower energy. The dependence
of $\lambda_{0}$ on $\alpha$ can be estimated using the following
intuitive argument. For a given $J_{z}<0$, the energy density of
the ferromagnetic state in the thermodynamic limit is given by $\epsilon_{{\rm FM}}=J_{z}\zeta(\alpha)$
{[}$\zeta(\alpha)\equiv\sum_{r=1}^{\infty}r^{-\alpha}$ is the Riemann
zeta function{]}, which diverges as $\alpha\rightarrow1$. For $J_{xy}=1$,
the magnitude of the energy density arising from the term $\sum_{i>j}(S_{i}^{x}S_{j}^{x}+S_{i}^{y}S_{j}^{y})/(i-j)^{\alpha}$
can be at most $\eta(\alpha)\equiv\sum_{r=1}^{\infty}(-1)^{r-1}/r^{\alpha}$
(the Dirichlet eta function), with this value obtained for any state
that is Néel-ordered along some direction in the $x-y$ plane. The
competition between the energy of these two classical states gives
a critical point $J_{z}\approx-\eta(\alpha)/\zeta(\alpha)$, which
smoothly varies from $J_{z}=-1$ at $\alpha=\infty$ to $J_{z}=0$
at $\alpha=1$. For $J_{xy}=-1$, the situation is quite different,
because the polarized state along any direction in the $x-y$ plane
has an energy density equal to $-\zeta(\alpha)$, and thus we naively
expect the phase boundary to be at $J_{z}=-1$ for all $\alpha>0$.

More formally, the boundary can be calculated via a spin-wave analysis.
We treat the spin state that is polarized along the $+z$ direction
as the vacuum state with no excitations, and apply the Holstein-Primakoff
transformation (for spin $1$) to map spin excitations (spin-waves)
into bosons: $S_{i}^{z}=1-a_{i}^{\dagger}a_{i}$, $S_{i}^{+}\equiv S_{i}^{x}+iS_{i}^{y}=\sqrt{2}a_{i}^{\dagger}(1-a_{i}^{\dagger}a_{i}/2)^{1/2}$.
In the weak excitation limit, $\langle a_{i}^{\dagger}a_{i}\rangle\ll1$,
we can approximate $S_{i}^{+}\approx\sqrt{2}a_{i}^{\dagger}$, and
our Hamiltonian becomes 
\begin{eqnarray}
H_{\text{sw}} & \approx & \sum_{i>j}\frac{-J_{z}(a_{i}^{\dagger}a_{i}+a_{j}^{\dagger}a_{j})+J_{xy}(a_{i}^{\dagger}a_{j}+a_{j}^{\dagger}a_{i})}{(i-j)^{\alpha}},\label{eq:Hsw}
\end{eqnarray}
where we have ignored the interaction terms $a_{i}^{\dagger}a_{i}a_{j}^{\dagger}a_{j}$
since $\langle a_{i}^{\dagger}a_{i}\rangle,\langle a_{j}^{\dagger}a_{j}\rangle\ll1$
is assumed. Assuming for the moment periodic boundary conditions,
this quadratic Hamiltonian can be diagonalized by Fourier transformation,
$H_{\text{sw}}=2\sum_{k}\omega_{k}c_{k}^{\dagger}c_{k}$, with the
following dispersion relation ($q\equiv2\pi k/N$) for an infinite
system

\begin{equation}
\omega(q)=-J_{z}\sum_{r=1}^{\infty}r^{-\alpha}+J_{xy}\sum_{r=1}^{\infty}\cos(qr)/r^{\alpha}.
\end{equation}

If $\omega_{min}\equiv\min\omega(q)>0$, then the ground state of
$H_{\text{sw}}$ is the vacuum state of all modes $k$, and $\langle a_{i}^{\dagger}a_{i}\rangle=0$
for all $i$, consistent with the approximation $\langle a_{i}^{\dagger}a_{i}\rangle\ll1$.
If $\omega_{min}<0$, then the ground state has an extensive number
of spin excitations and the spin-wave approximation should break down,
and we do not expect the polarized state in the $z$ direction to
be the quantum ground state. The $\omega_{min}=0$ condition thus
sets the phase boundary for $H_{\text{sw}}$. For $J_{xy}=1$, $\omega_{min}=\omega(q=\pi)=-J_{z}\zeta(\alpha)-\eta(\alpha)$,
leading to a critical line of $J_{z}=-\eta(\alpha)/\zeta(\alpha)$.
For $J_{xy}=-1$, $\omega_{min}=\omega(q=0)=(1-J_{z})\zeta(\alpha)$,
leading to a critical line at $J_{z}=-1$, independent of $\alpha$.
These results exactly match with the previous intuitive arguments.

\begin{figure}
\includegraphics[width=0.45\textwidth]{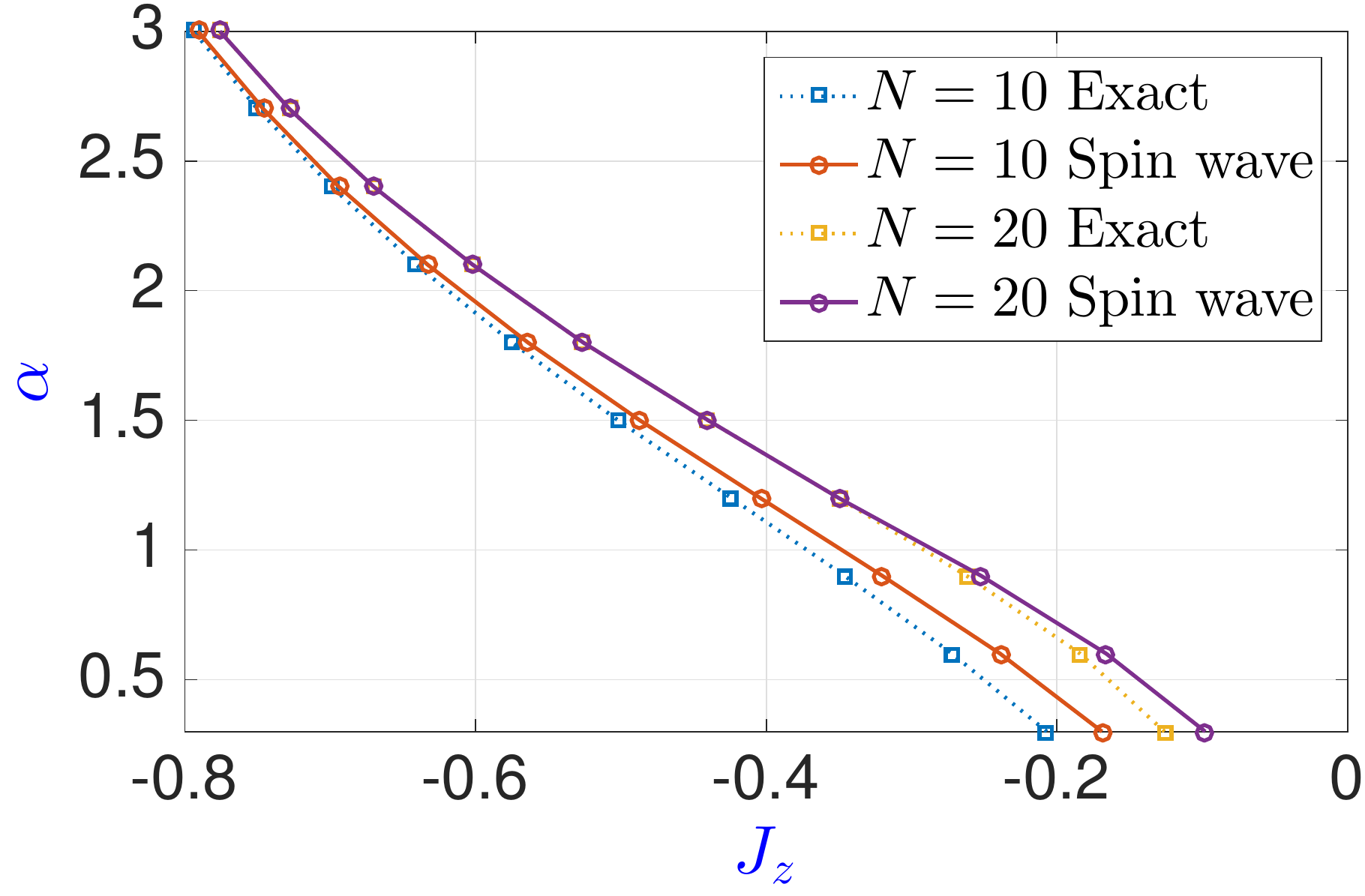}

\caption{\label{fig:EDspinwave} Comparison of the (first-order) transition
point out of the FM phase calculated using finite-size DMRG and spin-wave
theory for $J_{xy}=1$. The DMRG result is regarded as exact since
its error is far below the resolution of the plot.}
\end{figure}

Note that we can estimate the transition point of a finite chain without
translational invariance by numerically diagonalizing $H_{\text{sw}}$.
However, we note that the spin-wave analysis is not necessarily exact
in this case. The interactions between bosons that we have ignored
can make multi-particle eigenstates have a lower energy than the vacuum
state, despite the fact that the single-particle excitation spectrum
has a finite-size gap. In other words, the condition $\omega_{min}>0$
only guarantees the ferromagnetic state to be the ground state of
the non-interacting Hamiltonian $H_{\text{sw}}$ {[}Eq.\,\eqref{eq:Hsw}{]},
but not of the original Hamiltonian {[}Eq.\,\eqref{eq:H}{]}. This
effect of interactions can indeed be observed for finite-size systems.
In Fig.\,\ref{fig:EDspinwave}, we show that for $J_{xy}=1$, the
critical $J_{z}$ obtained by exact numerical calculations of Eq.\,\eqref{eq:H}
with $N=10$ spins is slightly smaller than spin-wave prediction given
by the condition $\omega_{min}=0$ for $0<\alpha<\infty$. To the
contrary, for $J_{xy}=-1$ we find that the spin-wave prediction is
exact for any number of spins and for all $\alpha>0$.

It is also interesting to note that the deviation of the transition
point due to the spin-wave approximation decreases with increasing
$\alpha$, and vanishes in the $\alpha\rightarrow\infty$ limit, showing
that long-range interactions are playing an important role. However,
using a finite-size DMRG algorithm \cite{schollwock_densitymatrix_2011,crosswhite_applying_2008,wall_outofequilibrium_2012,_dmrg},
we find that as $N$ increases, the deviation caused by the spin-wave
approximation decreases quickly (Fig.\,\ref{fig:EDspinwave}). In
addition, by using an infinite-size DMRG algorithm \cite{mcculloch_infinite_2008,_dmrg},
we find that the spin-wave prediction of the transition line $J_{z}=-\eta(\alpha)/\zeta(\alpha)$
{[}the green line in Fig.\,\ref{fig:iDMRG-SZZ}(a){]} is correct
within our numerical precision, strongly suggesting that the spin-wave
prediction becomes exact in the thermodynamic limit.

\section{Haldane phase and its boundary}

\begin{figure}
\includegraphics[width=0.45\textwidth]{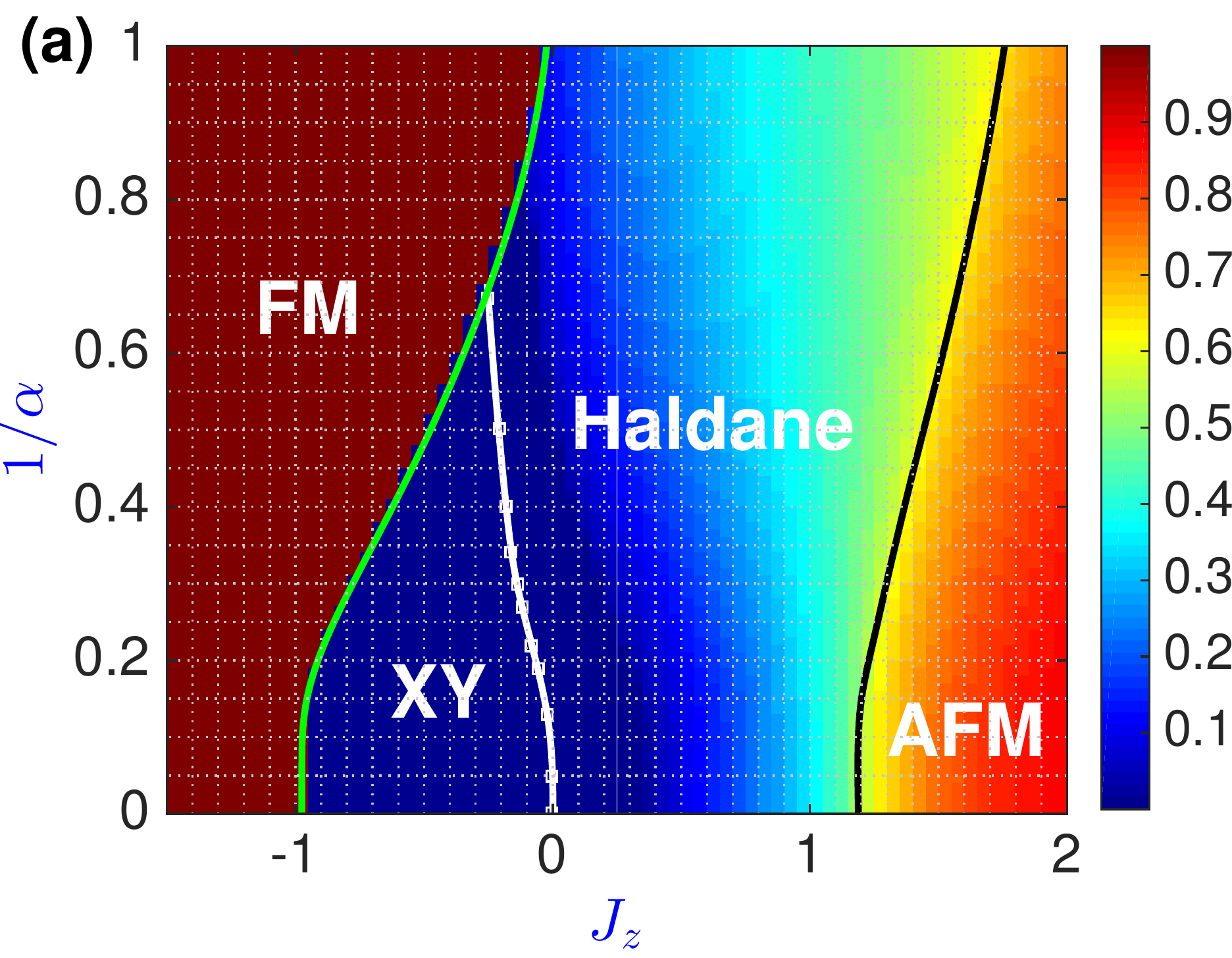}

\includegraphics[width=0.45\textwidth]{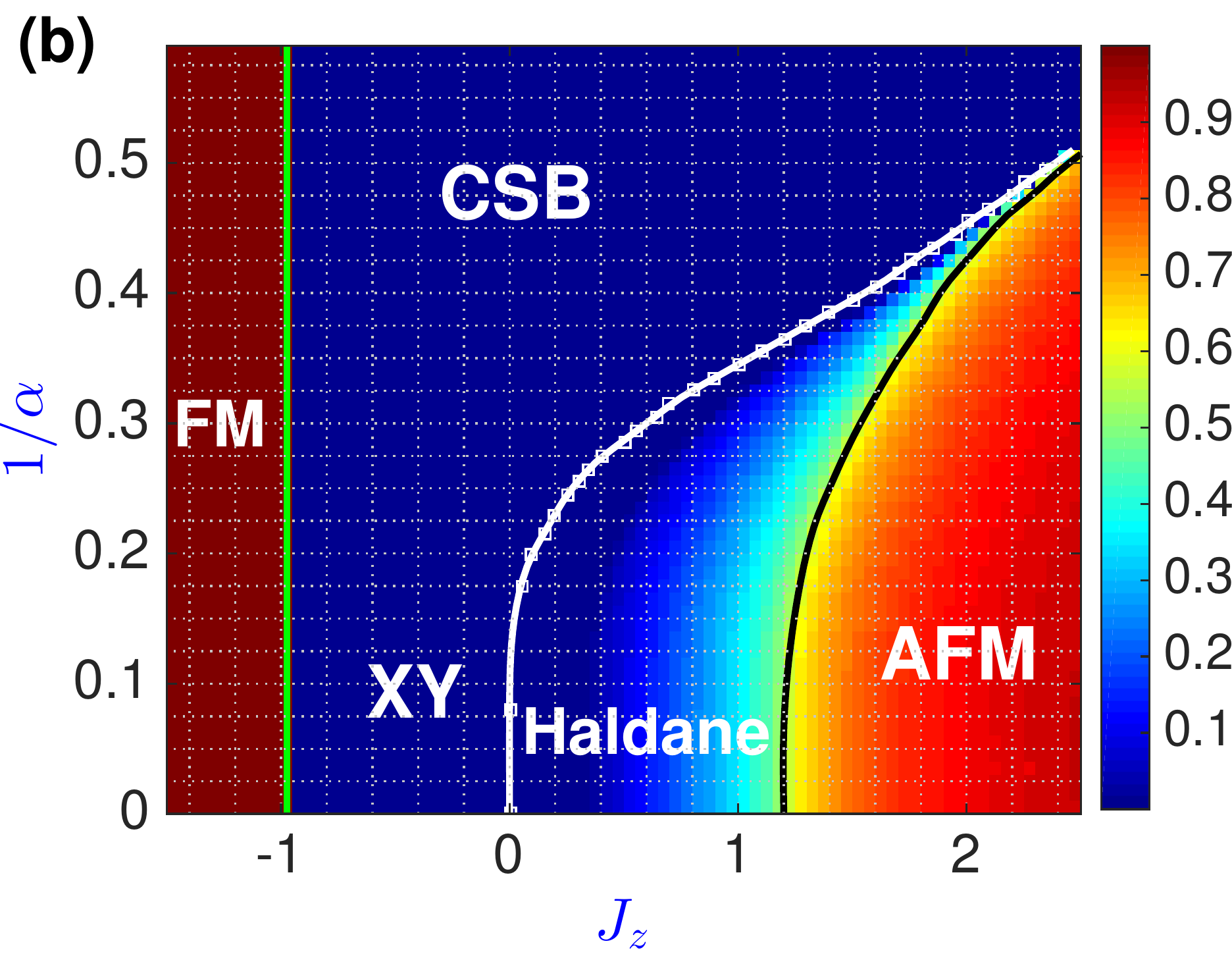}

\caption{\label{fig:iDMRG-SZZ}Infinite-size DMRG calculation of $\mathcal{S}_{ij}^{z}\equiv\langle S_{i}^{z}S_{j}^{z}\prod_{i<k<j}(-1)^{S_{k}^{z}}\rangle$
for a separation of $|i-j|=500$. $\mathcal{S}_{ij}^{z}=1$ in the
FM phase and $\mathcal{S}_{ij}^{z}\approx1$ deep in the AFM phase
for any $i$ and $j$. As $|i-j|\rightarrow\infty$, $\mathcal{S}_{ij}^{z}$
is finite for the Haldane phase and zero for the XY phase, thus we
can use it to locate the XY-to-Haldane phase boundary. (a) $J_{xy}=1$.
The FM phase boundary (green line) is given by the spin-wave prediction
$J_{z}=-\eta(\alpha)/\zeta(\alpha)$. (b) $J_{xy}=-1$. The FM phase
boundary (green line) is exactly at $J_{z}=-1$. For both (a) and
(b), we vary the bound dimension $\chi$ to accurately determine the
XY-to-Haldane phase boundary, determining the value of $J_{z}$ at
which $\mathcal{S}_{ij}^{z}$ vanishes (for a large but finite $|i-j|$)
and then extrapolating to the $\chi\rightarrow\infty$ limit (white
squares fitted by the white line). The black line is the Haldane-to-AFM
phase boundary, which is determined from $\langle S_{i}^{z}S_{j}^{z}\rangle$
(see text) and which converges well within the resolution of the plot
for $\chi\ge100$.}
\end{figure}

The existence of the Haldane phase in a spin-1 XXZ chain makes the
phase diagram much richer than that of a spin-1/2 XXZ chain. We focus
first on the XY-to-Haldane phase boundary $\lambda_{1}(\alpha)$.
The transition out of the Haldane phase is signaled by a vanishing
of the string-order correlation function $\mathcal{S}_{ij}^{\xi}\equiv\langle S_{i}^{\xi}S_{j}^{\xi}\prod_{i<k<j}(-1)^{S_{k}^{\xi}}\rangle$
($\xi=x,y,z$) when $|i-j|\rightarrow\infty$. However, because the
phase transition is of the BKT type, $\mathcal{S}_{ij}^{\xi}$ changes
rather smoothly with $J_{z}$ and $\alpha$ for a finite $|i-j|$,
and it is very challenging to find the exact transition point numerically.
Finite-size scaling using exact diagonalization on small chains must
be performed very carefully due to logarithmic corrections in system
size \cite{malvezzi_phase_1995,singh_ordering_1988,alcaraz_string_1992,kitazawa_phase_1996},
and infinite-size DMRG yields a phase transition point that depends
strongly on the bond dimension $\chi$ (the dimension of the matrix
product states used \cite{schollwock_densitymatrix_2011}), since
the ground state bipartite entanglement entropy $S$ grows logarithmically
with system size $N$ according to CFT: $S=c\log N+\text{const}$
\cite{calabrese_entanglement_2004}. As seen in Fig.\,\ref{fig:iDMRG-SZZ},
for $\chi=100$ and at $\alpha=\infty$, the string-order correlation
function $\mathcal{S}_{ij}^{z}$ appears to start vanishing at $J_{z}\approx0.3$,
instead of at $J_{z}=0$ as predicted by field theory \cite{schulz_phase_1986}.
However, this is consistent with previous infinite-size DMRG calculation
results \cite{su_nonlocal_2012,liu_entanglement_2014}. To extract
a more accurate phase boundary, we perform a scaling of $\chi$ ranging
from $50$ to $200$ near the XY-to-Haldane phase boundary, following
a procedure similar to that in Ref.\,\cite{su_nonlocal_2012}. We
then extract the XY-to-Haldane phase boundary (white line in Fig.\,\ref{fig:iDMRG-SZZ})
by determining the location where $\mathcal{S}_{ij}^{z}(\chi\rightarrow\infty)$
vanishes, which now correctly yields $J_{z}\approx0$ at $\alpha=\infty$.
However, we expect a few percent uncertainty in the transition point
due the use of $\mathcal{S}_{ij}^{z}$ at a finite separation $|i-j|$,
and due to the error in extrapolating $\mathcal{S}_{ij}^{z}(\chi\rightarrow\infty)$.

To explain why long-range interactions bend the XY-to-Haldane phase
boundary in opposite directions for ferromagnetic and antiferromagnetic
$J_{xy}$, we use an effective field theory first proposed by Haldane
\cite{haldane_continuum_1983} and developed by Affleck \cite{affleck_largen_1985}.
The proper inclusion of long-range interactions within this field
theoretic approach was discussed in detail in Ref.\,\cite{gong_topological_2015}.
Here, we give a brief review of this field-theory treatment. Consider
first the case of $J_{xy}=J_{z}=1$. In this case, each spin-1 is
mapped to a staggered field $\bm{n}(2i+\frac{1}{2})=(\bm{S}_{2i}-\bm{S}_{2i+1})/2$
and a uniform field $\bm{l}(2i+\frac{1}{2})=(\bm{S}_{2i}+\bm{S}_{2i+1})/2$.
Importantly, we observe that the classical ground state of $H$ is
always Néel-ordered for any $\alpha>0$, with $\bm{n}^{2}(x)=1$ and
$\bm{l}(x)=0$ for any position $x$. The intuition behind this decomposition
is that, in the quantum ground state, $\bm{n}(x)$ should have only
long-wave-length variations with $\bm{n}^{2}(x)\approx1$, while $\bm{l}(x)\approx0$
should represent long wave-length perturbations to the direction of
$\bm{n}(x)$ due to quantum fluctuations. Therefore, when working
with the Fourier-transformed fields $\bm{n}(q)$ and $\bm{l}(q)$,
we can expand the Hamiltonian in powers of the momentum $q$ and keep
only the leading order terms.

The effective Hamiltonian in the continuum limit and momentum space
reads 
\begin{equation}
H_{{\rm eff}}\approx\int dq\left[\omega(q)|\bm{n}(q)|^{2}+\Omega(q)|\bm{l}(q)|^{2}\right],\label{eq:Heff}
\end{equation}
where the cross terms between $\bm{n}$ and $\bm{l}$ are ignored
because they involve $\bm{n}(q)$ near $q=\pi$. The dispersion relations
$\Omega(q)$ and $\omega(q)$ can be expanded at small $q$ as \cite{NonIntegerAlpha}:
\begin{eqnarray}
\omega(q) & \equiv & 2\sum_{r=1}^{\infty}(-1)^{r}\frac{\cos(qr)}{r^{\alpha}}\approx-2\eta(\alpha)+\eta(\alpha-2)q^{2}+O(q^{4}),\nonumber \\
\Omega(q) & \equiv & 2\sum_{r=1}^{\infty}\frac{\cos(qr)}{r^{\alpha}}\approx2\zeta(\alpha)+\zeta(\alpha-2)q^{2}+O(q^{4})\nonumber \\
 &  & +2\Gamma(1-\alpha)\cos[\frac{\pi}{2}(\alpha-1)]|q|^{\alpha-1}.\label{eq:Oq}
\end{eqnarray}

For the $\bm{n}$ field, we need to keep the $q^{2}$ term since the
zeroth-order term gives a constant due to the approximation $\bm{n}^{2}(x)\approx1$.
The zeroth-order term in $q$ for the $\bm{l}$ field is the dominant
source of quantum fluctuations, and we can ignore higher-order terms
in determining whether $H_{{\rm eff}}$ is gapped or not (they do
contribute to the long-distance behavior of correlation functions
though \cite{gong_topological_2015}). Because $H_{{\rm eff}}$ is
quadratic in both fields, we can first integrate out the $\bm{l}$
field using the standard coherent spin-state path integral \cite{sachdev_quantum_2011}.
We then obtain a 1D O$(3)$ nonlinear sigma model (NLSM) of the field
$\bm{n}$ \cite{haldane_continuum_1983}, which can be treated by
removing the nonlinear constraint $\bm{n}^{2}=1$ while phenomenologically
introducing a gap $\Delta_{\alpha}$ and a renormalized spin-wave
velocity $v_{\alpha}$ \cite{sorensen_sk_1994,gong_topological_2015}.
We thereby arrive at a free field theory with the Lagrangian density
(written in momentum space) 
\begin{equation}
\mathcal{L}(q)\propto\left(\frac{\partial\bm{n}}{\partial t}\right)^{2}-(\Delta_{\alpha}^{2}+v_{\alpha}^{2}q^{2})|\bm{n}(q)|^{2}.\label{eq:FreeL}
\end{equation}
The existence of the gap can be understood by the renormalization
group flow of the coupling strength \cite{sachdev_quantum_2011,fradkin_field_2013},
or by considering the SU$(n)$ variant of the Hamiltonian, for which
the corresponding O$(n)$ NLSM can be solved analytically in the $n\rightarrow\infty$
limit and give rise to a mass gap \cite{affleck_largen_1985,brezin_fields_1990}.
We infer that $\Delta_{\alpha}$ should increase as $\alpha$ decreases,
since $\Delta_{\alpha\rightarrow\infty}\approx0.41$ \cite{white_density_1992,white_numerical_1993}
and $\Delta_{a\rightarrow0}=1$ (where the Hamiltonian becomes integrable).
This speculation is confirmed by accurate finite-size DMRG calculation
of $\Delta_{\alpha}$ \cite{gong_topological_2015}.

Next, we consider the case of $J_{xy}=1$ but $J_{z}<1$. We can then
write 
\begin{equation}
H=\sum_{i>j}\frac{1}{(i-j)^{\alpha}}\bm{S}_{i}\cdot\bm{S}_{j}-(1-J_{z})\sum_{i>j}\frac{1}{(i-j)^{\alpha}}S_{i}^{z}S_{j}^{z}.\label{eq:Hani}
\end{equation}
Following Refs.\,\cite{affleck_largen_1985,murashima_phase_2005},
the anisotropy term above can be treated as a negative mass term $(1-J_{z})f_{\alpha}n_{z}^{2}(q)$
to the Lagrangian density $\mathcal{L}(q)$ in Eq.\,\eqref{eq:FreeL}.
The precise value of the renormalization factor $f_{\alpha}$ is not
important to us, but we expect it to continuously decrease as $\alpha$
becomes smaller, since the staggered field dominates in the Haldane
phase and long-range interactions {[}$\sum_{i>j}\frac{1}{(i-j)^{\alpha}}S_{i}^{z}S_{j}^{z}$
in Eq.\,\eqref{eq:Hani}{]} are increasingly frustrated as $\alpha$
decreases. The mass gap for the field $n_{z}$ is now smaller than
for $n_{x}$ and $n_{y}$, and reads $\Delta_{\alpha}(J_{z})=\sqrt{\Delta_{\alpha}^{2}-(1-J_{z})f_{\alpha}}$.
Combined with the above discussion that $\Delta_{\alpha}$ should
increase with decreasing $\alpha$, we require progressively more
negative $J_{z}$ to close the gap and transition into the XY phase
as $\alpha$ decreases, thus explaining the shape of the XY-to-Haldane
phase boundary in Fig.\,\ref{fig:iDMRG-SZZ}(a).

For $J_{xy}=-1$ and $J_{z}<1$, the classical ground state is no
longer Néel ordered and the field theory employed above is not valid.
However, by rotating every other spin by $\pi$ about the $z$-axis,
we generate a transformed Hamiltonian 
\begin{equation}
H^{\prime}=\sum_{i>j}\frac{(-1)^{i-j-1}}{(i-j)^{\alpha}}\bm{S}_{i}\cdot\bm{S}_{j}+\sum_{i>j}\frac{J_{z}-(-1)^{i-j-1}}{(i-j)^{\alpha}}S_{i}^{z}S_{j}^{z}.\label{eq:H'}
\end{equation}
Now the classical ground state \emph{is} Néel ordered (along any direction
for $J_{z}=1$). The first term above is isotropic, and gets mapped
to 
\begin{equation}
\sum_{i>j}\frac{(-1)^{i-j-1}}{(i-j)^{\alpha}}\bm{S}_{i}\cdot\bm{S}_{j}\approx\int dq\left[\Omega(q)|\bm{n}(q)|^{2}+\omega(q)|\bm{l}(q)|^{2}\right],\label{eq:Hstag}
\end{equation}
where the roles of $\omega(q)$ and $\Omega(q)$ are swapped as compared
to Eq.\,\eqref{eq:Heff}. For $\alpha<3$, $\Omega(q)$ in Eq.\,\eqref{eq:Oq}
is now dominated by the non-analytic term $|q|^{\alpha-1}$ at small
$q$, and we can no longer obtain the simple free Lagrangian in Eq.\,\eqref{eq:FreeL}.
In Ref.\,\cite{gong_topological_2015}, it is shown that the $|q|^{\alpha-1}$
term in the dispersion of $\bm{n}(q)$ in Eq.\,\eqref{eq:Hstag}
leads to a renormalization group flow towards a gapless ordered phase
spontaneously breaking an $SU(2)$ symmetry for $\alpha<\alpha_{c}\lesssim3$.
For our complete Hamiltonian $H^{\prime}$ in Eq.\,\eqref{eq:H'},
the anisotropy leads instead to a $U(1)$ continuous symmetry breaking
phase for $\alpha<\alpha_{c}^{\prime}$ (see the next section for
further discussions, where $\alpha_{c}^{\prime}$ is estimated to
be $2.9$ at $J_{z}=1$). Our infinite-size DMRG calculations in Fig.\,\ref{fig:iDMRG-SZZ}(b)
suggest that the Haldane phase terminates at a critical $\alpha$
around $3.1$ for $J_{z}=1$, and the XY phase is expected to exist
in between the CSB phase and the Haldane phase at $J_{z}=1$.

For $\alpha>3$, $\Omega(q)$ is dominated by $q^{2}$ and we can
once again reduce $H^{\prime}$ to the free field Lagrangian Eq.\,\eqref{eq:FreeL},
but with a different mass gap $\Delta_{\alpha}^{\prime}$ and spin-wave
velocity $v_{\alpha}^{\prime}$. The anisotropy term in Eq.\,\eqref{eq:H'}
changes the gap to $\Delta_{\alpha}^{\prime}(J_{z})=\sqrt{\Delta_{\alpha}^{\prime2}-(g_{\alpha}-J_{z}h_{\alpha})}$.
Here $g_{\alpha}$ is a renormalization factor due to non-frustrating
long-range interactions $\frac{(-1)^{i-j-1}}{(i-j)^{\alpha}}S_{i}^{z}S_{j}^{z}$
in Eq.\,\eqref{eq:H'}, and should thus increase as $\alpha$ decreases,
while $h_{\alpha}$ is a renormalization factor due to frustrating
long-range interaction $\frac{1}{(i-j)^{\alpha}}S_{i}^{z}S_{j}^{z}$
in Eq.\,\eqref{eq:H'}, and should decrease as $\alpha$ decreases.
Together with the expectation that the gap $\Delta_{\alpha}^{\prime}$
should decrease with $\alpha$ \cite{gong_topological_2015} due to
the appearance of gapless continuous symmetry breaking phase at $\alpha\lesssim3$,
we conclude that the gap closes at a point with $J_{z}$ strictly
larger than zero in the presence of long-range interactions, again
consistent with our numerical results.

We point out that a different field theoretic approach based on non-Abelian
bosonization \cite{schulz_phase_1986,gong_topological_2015} can also
be employed to predict the qualitative changes to the XY-to-Haldane
phase boundary. This technique has been used to predict the XY-to-Haldane
phase boundary of a spin-1 XXZ chain with next-nearest-neighbor interactions
\cite{murashima_phase_2005}, which is a reasonable approximation
to our model when $\alpha$ is large enough that next-nearest-neighbor
interactions dominate over the next-longer-range interactions.

We end this section with a brief discussion of the boundary between
the Haldane and AFM phases. Both the Haldane and AFM phases are gapped
and have finite entanglement entropy in the infinite-system-size limit
\cite{arealaw}. Thus our infinite-size DMRG calculations should precisely
reproduce the phase boundary between the two phases; indeed we see
well-converged results for bond dimensions of $\chi\ge100$. We extract
the Haldane-to-AFM phase boundaries using the spin-spin correlation
functions $C_{ij}^{z}\equiv\langle S_{i}^{z}S_{j}^{z}\rangle$ (not
shown), and plot them as black lines in Figs.\,\ref{fig:iDMRG-SZZ}(a,b).
The bending of the Haldane-to-AFM phase boundary toward larger $J_{z}$
for both $J_{xy}=1$ and $J_{xy}=-1$ in the presence of long-range
interactions can be understood via simple energetic considerations.
In the AFM phase, the spins are (nearly) anti-aligned in the $z$
direction; long-range interactions are strongly frustrated, and the
energy $E=\sum_{i>j}\langle S_{i}^{z}S_{j}^{z}\rangle/(i-j)^{\alpha}$
at $\alpha\rightarrow0$ is only half of the $\alpha=\infty$ case
for a perfectly Néel ordered state. In the Haldane phase, the AFM
order of spin correlations $\langle\bm{S}_{i}\cdot\bm{S}_{j}\rangle$
decays exponentially (followed by a small power-law tail \cite{gong_topological_2015}),
and thus the ground state energy $E=\sum_{i>j}\langle\bm{S}_{i}\cdot\bm{S}_{j}\rangle/(i-j)^{\alpha}$
is much less frustrated by the long-range interactions. As a result,
we expect the disordered ground state in the Haldane phase to have
progressively lower energy than an AFM ordered state as $\alpha$
decreases at a given $J_{z}$, and hence a larger (but always finite
even for $\alpha\rightarrow0$) $J_{z}$ is needed to make the transition
from the Haldane phase into the AFM phase.

\section{CSB Phase and its boundary}

The celebrated Mermin-Wagner theorem rigorously rules out continuous
symmetry breaking in 1D and 2D quantum and classical spin systems
at finite temperature, as long as the interactions satisfy the convergence
condition $\sum_{i>j}J_{ij}r_{ij}^{2}<\infty$ in the thermodynamic
limit ($r_{ij}$ and $J_{ij}$ are respectively the distance and coupling
strength between sites $i$ and $j$) \cite{mermin_absence_1966}.
The long-distance properties of 1D spin systems at zero temperature
can often be related to those of a 2D classical model at finite temperature;
however, in the process of this mapping, the long-range interactions
are only inherited by one of the two spatial directions in the classical
model, and the Mermin-Wagner convergence condition will be satisfied
for interactions decaying faster than $1/r^{3}$. Thus we expect no
continuous symmetry breaking in the ground state of our Hamiltonian
Eq.\,\eqref{eq:H} for $\alpha>3$. Indeed, we have found exclusively
disordered or discrete ($Z_{2}$) symmetry breaking phases for $\alpha>3$
in our phase diagrams (Fig.\,\ref{fig:PhaseDiagram}). Continuous
symmetry breaking can (and does) appear when $\alpha<3$ . To gain
a better understanding of the robustness of symmetry breaking states
to quantum fluctuations, below we carry out a spin-wave analysis \cite{auerbach_interacting_1994}.

We start by considering the $J_{xy}=-1$ case, and take the state
with all spins polarized along the $+x$ direction as the vacuum state.
With this choice of vacuum, and assuming that the density of spin
waves is small ($\langle a_{i}^{\dagger}a_{i}\rangle\ll1$ in the
following expressions), the Holstein-Primakoff mapping is now $S_{i}^{x}=1-a_{i}^{\dagger}a_{i}$,
$S_{i}^{y}\approx(a_{i}^{\dagger}+a_{i})/\sqrt{2}$, $S_{i}^{z}\approx(a_{i}^{\dagger}-a_{i})/i\sqrt{2}$.
Under this mapping, and dropping terms that are quartic in bosonic
operators (again based on the assumption that $\langle a_{i}^{\dagger}a_{i}\rangle\ll1$),
$H$ becomes 
\begin{eqnarray}
H_{\text{swx}} & = & \sum_{k=-N/2}^{N/2}\begin{pmatrix}a_{k}^{\dagger} & a_{-k}\end{pmatrix}\begin{pmatrix}\omega_{k} & \mu_{k}\\
\mu_{k} & \omega_{k}
\end{pmatrix}\begin{pmatrix}a_{k}\\
a_{-k}^{\dagger}
\end{pmatrix};\label{eq:Hswx}\\
\omega_{k} & = & \sum_{r=1}^{N/2}J_{r}+\frac{J_{z}-1}{2}\sum_{r=1}^{N/2}J_{r}\cos(\frac{2\pi k}{N}r),\label{eq:wk}\\
\mu_{k} & = & -\frac{J_{z}+1}{2}\sum_{r=1}^{N/2}J_{r}\cos(\frac{2\pi k}{N}r),\label{eq:uk}
\end{eqnarray}
where $a_{k}=\frac{1}{\sqrt{N}}\sum_{j}e^{i2\pi jk/N}a_{j}$. $H_{\text{swx}}$
can be diagonalized with a Bogoliubov transformation, yielding non-interacting
Bogoliubov quasi-particles with a spectrum $\nu_{k}$. Importantly,
when $|\omega_{k}|>|\mu_{k}|$, $\nu_{k}>0$ and the vacuum is dynamically
stable. When $|\omega_{k}|<|\mu_{k}|$, $\nu_{k}$ is imaginary and
the system is dynamically unstable indicating that we have made the
wrong choice of a classical ground state. Using the expressions for
$\omega_{k}$ and $\mu_{k}$ in Eqs.\,\eqref{eq:wk} and \eqref{eq:uk},
we find that $|\omega_{k}|>|\mu_{k}|$ is satisfied for all $k\ne0$
modes if and only if $-1\le J_{z}<\zeta(\alpha)/\eta(\alpha)$. This
is because when $J_{z}<-1$, the classical ground state is ferromagnetic
in $z$ direction, and when $J_{z}>\zeta(\alpha)/\eta(\alpha)$ the
classical ground state is Néel ordered along the $z$ direction.

Because the Bogoliubov quasiparticles consist of both particles and
holes, the ground state of $H_{\text{swx}}$ can have a finite or
even divergent density of spin excitations, measured by 
\begin{eqnarray}
\langle a_{i}^{\dagger}a_{i}\rangle & = & \frac{1}{N}\sum_{k\ne0}\frac{1}{2}([1-\mu_{k}^{2}/\omega_{k}^{2}]^{-1/2}-1)\label{eq:density}\\
 & \xrightarrow{N\rightarrow\infty} & \frac{1}{4\pi}\int_{-\pi}^{\pi}dq\left([1-\mu^{2}(q)/\omega^{2}(q)]^{-1/2}-1\right).\nonumber 
\end{eqnarray}

\begin{figure}
\includegraphics[width=0.45\textwidth]{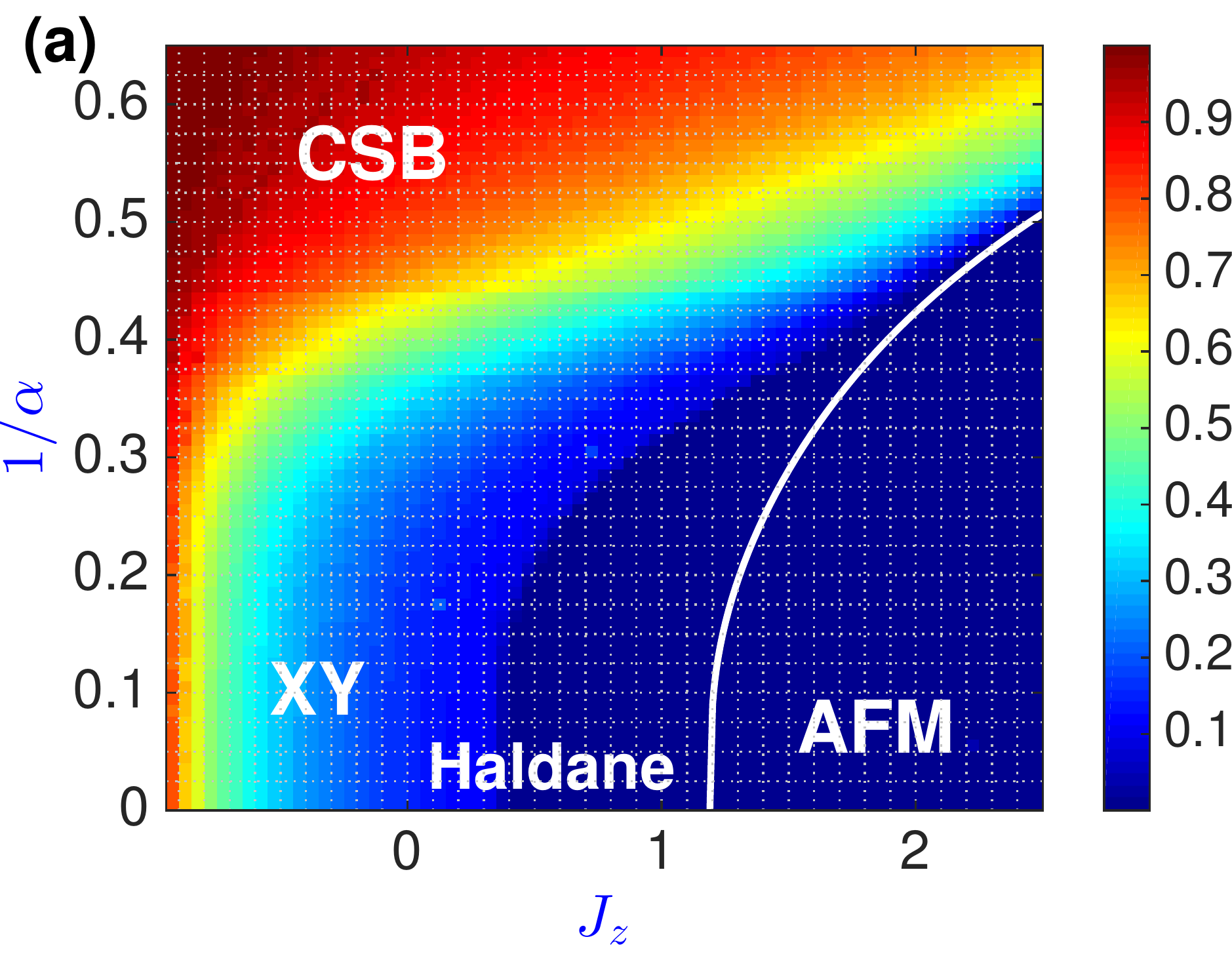}

\includegraphics[width=0.45\textwidth]{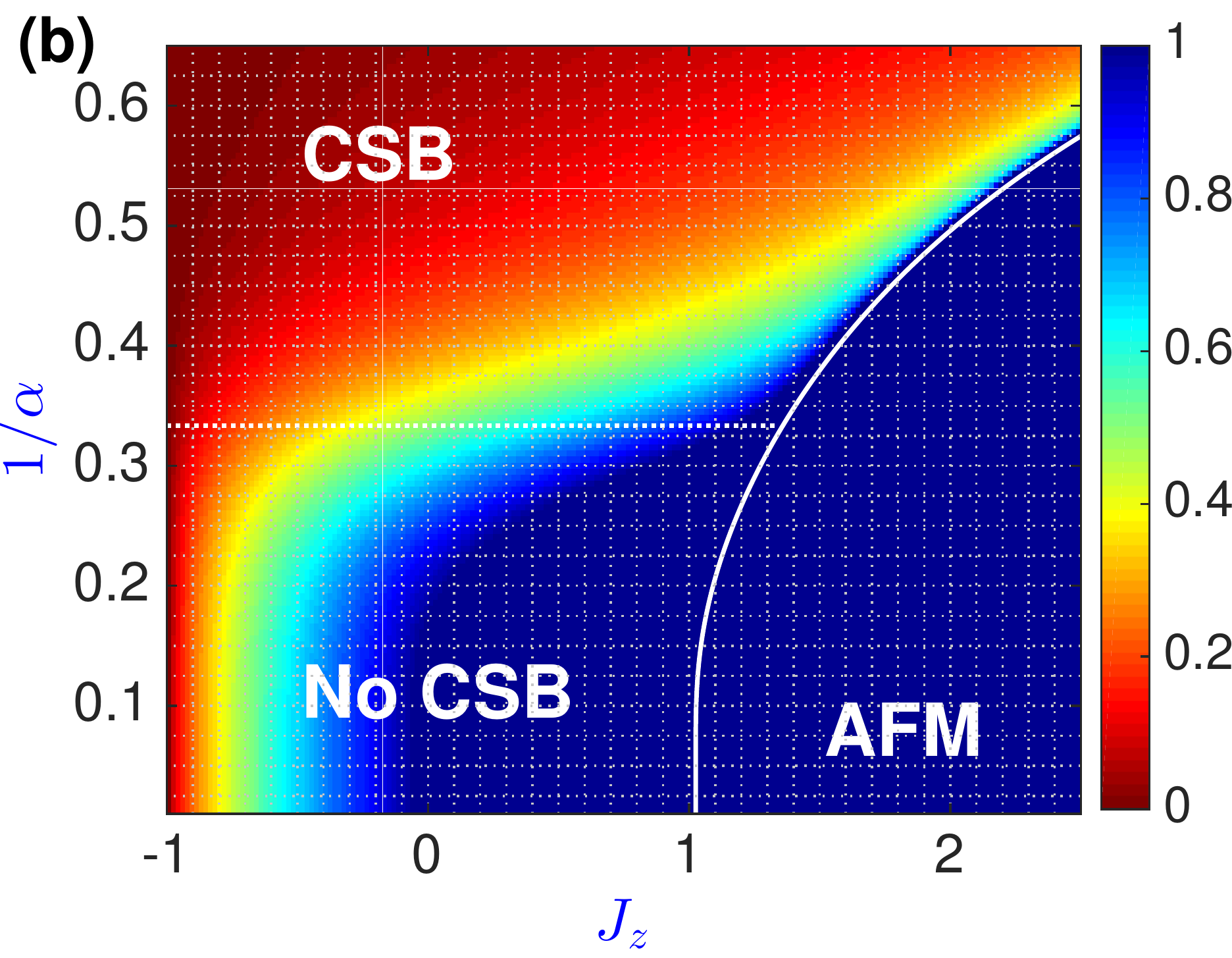}

\caption{\label{fig:CSB} Continuous symmetry breaking for $J_{xy}=-1$. (a)
iDMRG calculation of $\langle S_{i}^{+}S_{j}^{-}\rangle$ for $|i-j|=500$,
with bond dimension $\chi=200$ (at this separation and bond dimension,
the results are well converged). Long-range order in the $x-y$ plane
is increasingly favored as $\alpha$ decreases, but we can not extract
a sharp phase boundary between the CSB and XY phase because an impractically
large bond dimension is needed to accurately extract the simultaneous
$\chi,|i-j|\rightarrow\infty$ limit of $\langle S_{i}^{+}S_{j}^{-}\rangle$.
The white line denoting the boundary of the AFM phase is from Fig.\,\ref{fig:iDMRG-SZZ}(a).
(b) Spin-wave excitation density $\langle a_{i}^{\dagger}a_{i}\rangle$
calculated using Eq.\,\eqref{eq:density} for $N=1001$ spins. For
$J_{z}>\zeta(\alpha)/\eta(\alpha)$ imaginary frequencies appear in
the Bogoliubov spectrum, indicating a classical instability toward
the AFM phase. We set $\langle a_{i}^{\dagger}a_{i}\rangle=1$ in
this region, as well as in regions where $\langle a_{i}^{\dagger}a_{i}\rangle>1$.
For $\alpha>3$, $\langle a_{i}^{\dagger}a_{i}\rangle\rightarrow\infty$
as $N\rightarrow\infty$, thus no CSB phase is expected (boundary
shown by the white dashed line). }
\end{figure}

The integrand $[1-\mu^{2}(q)/\omega^{2}(q)]^{-1/2}$ above diverges
at $q=0$, and whether or not the integral is infrared divergent depends
on the value of $\alpha$. We find that for $\alpha>3$, $[1-\mu^{2}(q)/\omega^{2}(q)]^{-1/2}\propto|q|^{-1}$
to leading order in $q$, and therefore $\langle a_{i}^{\dagger}a_{i}\rangle\sim\ln(N)$
diverges as $N\rightarrow\infty$. This means the long-range ferromagnetic
order along the $x$ direction is destroyed by quantum fluctuations
in the thermodynamic limit; we expect that $\lim_{|i-j|\rightarrow\infty}\langle S_{i}^{+}S_{j}^{-}\rangle=0$,
and the system will be disordered (either Haldane or XY). For $\alpha<3$,
we find that $[1-\mu^{2}(q)/\omega^{2}(q)]^{-1/2}\propto|q|^{-(\alpha-1)/2}$
to leading order in $q$, and the integral is infrared convergent.
The excitation density $\langle a_{i}^{\dagger}a_{i}\rangle$ converges
to a finite constant, so we expect a CSB phase with $\lim_{|i-j|\rightarrow\infty}\langle S_{i}^{+}S_{j}^{-}\rangle\ne0$.
However, when $\langle a_{i}^{\dagger}a_{i}\rangle$ converges to
a constant on the order of $1$, the spin-wave approximation is not
expected to be accurate, and it is possible that the actual ground
state of $H$ remains disordered for $\alpha$ slightly less than
$3$.

For $J_{xy}=1$, classically the spins prefer to anti-align in the
$x-y$ plane. Expanding around this classical state with the same
spin-wave approximation, both $\mu(q)$ and $\omega(q)$ become fully
analytic due to an additional alternating sign $(-1)^{r}$ in Eqs.\,\eqref{eq:wk}
and \eqref{eq:uk}. As a result, $[1-\mu^{2}(q)/\omega^{2}(q)]^{-1/2}$
always exhibits a $|q|^{-1}$ divergence at small $q$, and continuous
symmetry breaking is forbidden for all $\alpha>0$.

From our infinite-size DMRG calculations, we see that $\langle S_{i}^{+}S_{j}^{-}\rangle\sim1/|i-j|^{\eta}$
decays with a rather slow power law in the XY phase (e.g. $\eta=0.25$
at $J_{z}=0$ and $\alpha=\infty$; $\eta$ is non-universal and depends
on $J_{z}$ and $\alpha$). At the maximum separation that we can
calculate accurately, $\langle S_{i}^{+}S_{j}^{-}\rangle$ only shows
a crossover from the XY phase to the CSB phase {[}Fig.\,\ref{fig:CSB}(a){]}.
This crossover can in fact be qualitatively reproduced using the above
spin-wave theory by calculating the spin-wave excitation density $\langle a_{i}^{\dagger}a_{i}\rangle$
for a finite system size {[}Fig.\,\ref{fig:CSB}(b){]}.

Further numerical evidence of the CSB phase is obtained by calculating
the effective central charge $c_{\text{eff}}$ as a function of $\alpha$
and $J_{z}$, which can be obtained by calculating the half-chain
entanglement entropy $S$ for two chains with different total lengths
$N_{1}$ and $N_{2}$ using a finite-size DMRG algorithm. Explicitly,
for large $N_{1}$ and $N_{2}$, we have 
\begin{equation}
c_{\text{eff}}\approx6\frac{S(N_{1})-S(N_{2})}{\ln(N_{1})-\ln(N_{2})}.\label{eq:ceff}
\end{equation}

In the XY phase (including its boundaries) and at the boundary between
the Haldane and AFM phases, we expect 1+1D conformal symmetry in the
underlying field theory model \cite{alcaraz_critical_1992,ejima_comparative_2015},
with $c_{\text{eff}}$ being the actual central charge representing
the conformal anomaly \cite{calabrese_entanglement_2004}. In the
Haldane, FM, and AFM phases, no 1+1D conformal symmetry exists due
to the presence of a gap. Although the CSB phase is gapless, we expect
a breakdown of 1+1D conformal symmetry due to the $1/r^{\alpha}$
long-range interactions that become relevant in the RG sense for $\alpha\lesssim3$
\cite{vodola_kitaev_2014,vodola_longrange_2015,maghrebi_continuous_2015}.
We emphasize that in phases with no conformal symmetry, $c_{\text{eff}}$
does not have the meaning of the central charge and is used only as
a diagnostic here to numerically find phase boundaries.

\begin{figure}[H]
\includegraphics[width=0.45\textwidth]{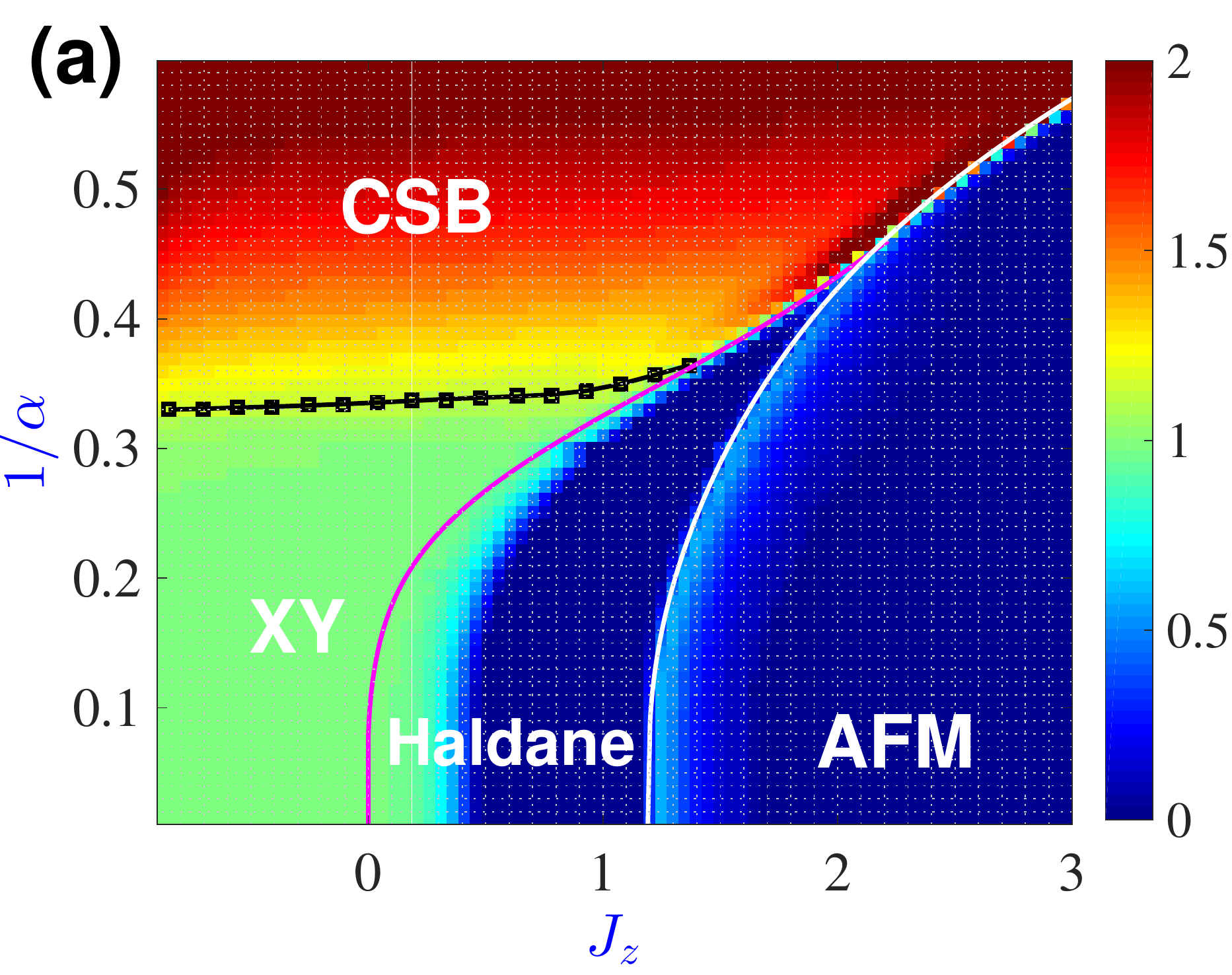}

\includegraphics[width=0.42\textwidth]{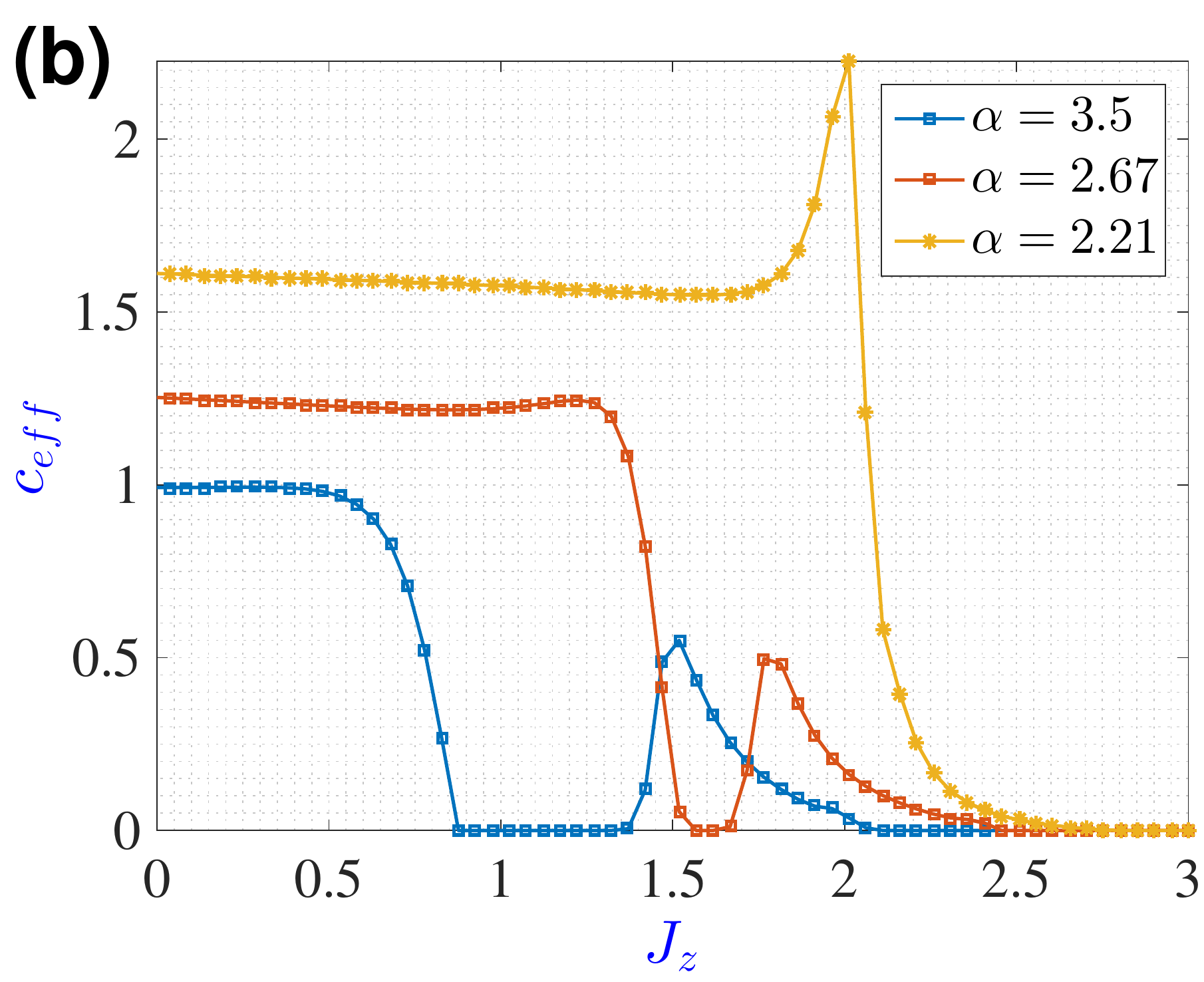}

\caption{\label{fig:DMRG-c}Calculation of the effective central charge $c_{\text{eff}}$
as a function of $J_{z}$ and $\alpha$ for $J_{xy}=-1$, extracted
from finite-size DMRG calculations with $N_{1}=100$, $N_{2}=110$,
and a maximum bond dimension of $500$. (a) The black squares (fitted
by the black line) show where $c_{\text{eff}}$ starts to deviate
from $1$ when going from the XY to the CSB phase. The purple line
and white line are from Fig.\,\ref{fig:iDMRG-SZZ}, and show the
boundaries of the Haldane phase. (The calculation of $c_{\text{eff}}$
is inaccurate in predicting the location of the XY-to-Haldane transition
due to strong finite-size effects \cite{malvezzi_phase_1995,singh_ordering_1988,alcaraz_string_1992,kitazawa_phase_1996}.)
For better contrast, locations with $c>2$ are shown with the color
corresponding to $c=2$. (b) For our finite-size chains, the XY-to-Haldane
BKT phase transition is signaled by a continuous drop of $c_{\text{eff}}$
from $1$ to $0$ ($\alpha=3.5$). The Haldane-to-AFM phase transition
is identified by a peak with value around $0.5$ in $c_{\text{eff}}$
($\alpha=3.5$ and $\alpha=2.67$). The CSB-to-Haldane transition
is expected to be continuous and not associated with a central charge
($\alpha=2.67$). The CSB-to-AFM transition has a sharp peak in $c_{\text{eff}}$
($\alpha=2.21$), an indication of a first-order transition \cite{ejima_comparative_2015}.}
\end{figure}

We identify the XY-to-CSB phase boundary in Fig.\,\ref{fig:DMRG-c}
as the place where $c_{\text{eff}}$ starts to become appreciably
(5-10\%) larger than $1$. Due to finite-size effects, $c_{\text{eff}}$
changes continuously for continuous phase transitions, and we are
not able to obtain the precise location of the XY-to-CSB phase boundary.
Nevertheless, we find good agreement with the XY-to-CSB phase boundary
predicted by spin-wave theory, especially near $J_{z}=-1$, where
spin-wave theory should be almost exact. Together with perturbative
field theory calculations presented in Ref.\,\cite{maghrebi_continuous_2015},
we expect the phase boundary in Fig.\,\ref{fig:DMRG-c} to be accurate
within a few percent.  A CSB-XY-Haldane tricritical point is found
at $\alpha\approx2.75$ and $J_{z}\approx1.35$.

From Ref.\,\cite{maghrebi_continuous_2015}, it follows that the
XY-to-CSB transition is a BKT-like transition that belongs to a universality
class different from the XY-to-Haldane BKT transition. The Haldane-to-CSB
transition is somewhat exotic, because the Haldane phase maps to a
high-temperature disordered phase in a 2D classical model \cite{fradkin_field_2013},
and \emph{in the absence of long-range interactions}, the CSB phase
exists in 2D only at zero temperature \cite{mermin_absence_1966}
and is unlikely to undergo a phase transition directly to a high-temperature
disordered phase. We also argue that the CSB-to-Haldane transition
is not described by a 1+1D CFT, as supported by our numerical calculations
shown in Fig.\,\ref{fig:DMRG-c}(b), where $c_{\text{eff}}$ changes
smoothly (at least for finite chains) from a value larger than $1$
to $0$ during the CSB-to-Haldane transition.

The CSB-to-AFM phase transition is very likely to be first-order,
similar to the transition between the large-$D$ and AFM phases studied
in Refs.\,\cite{chen_groundstate_2003,liu_entanglement_2014}, despite
the existence of quantum fluctuations in both phases. As shown in
Fig.\,\ref{fig:DMRG-c}, we observe a sharp peak in $c_{\text{eff}}$
at small $\alpha$s when $J_{z}$ is varied, indicating a first order
transition \cite{ejima_comparative_2015}, with further evidence that
includes jumps in sub-lattice magnetization and spin-spin correlation
across the CSB-to-AFM transition (not shown).

\section{Experimental Detection}

\begin{figure}[H]
\includegraphics[width=0.5\columnwidth]{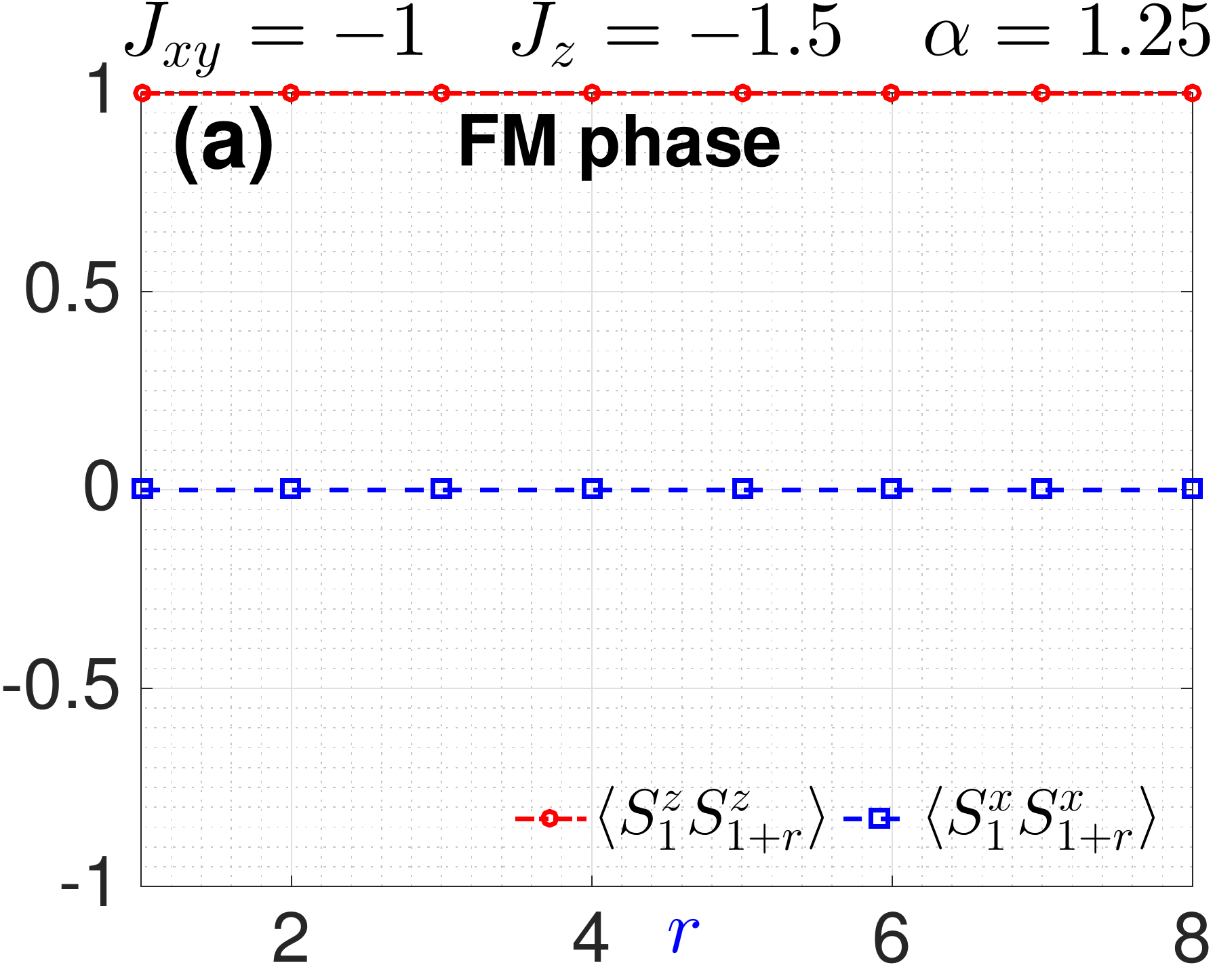}\includegraphics[width=0.5\columnwidth]{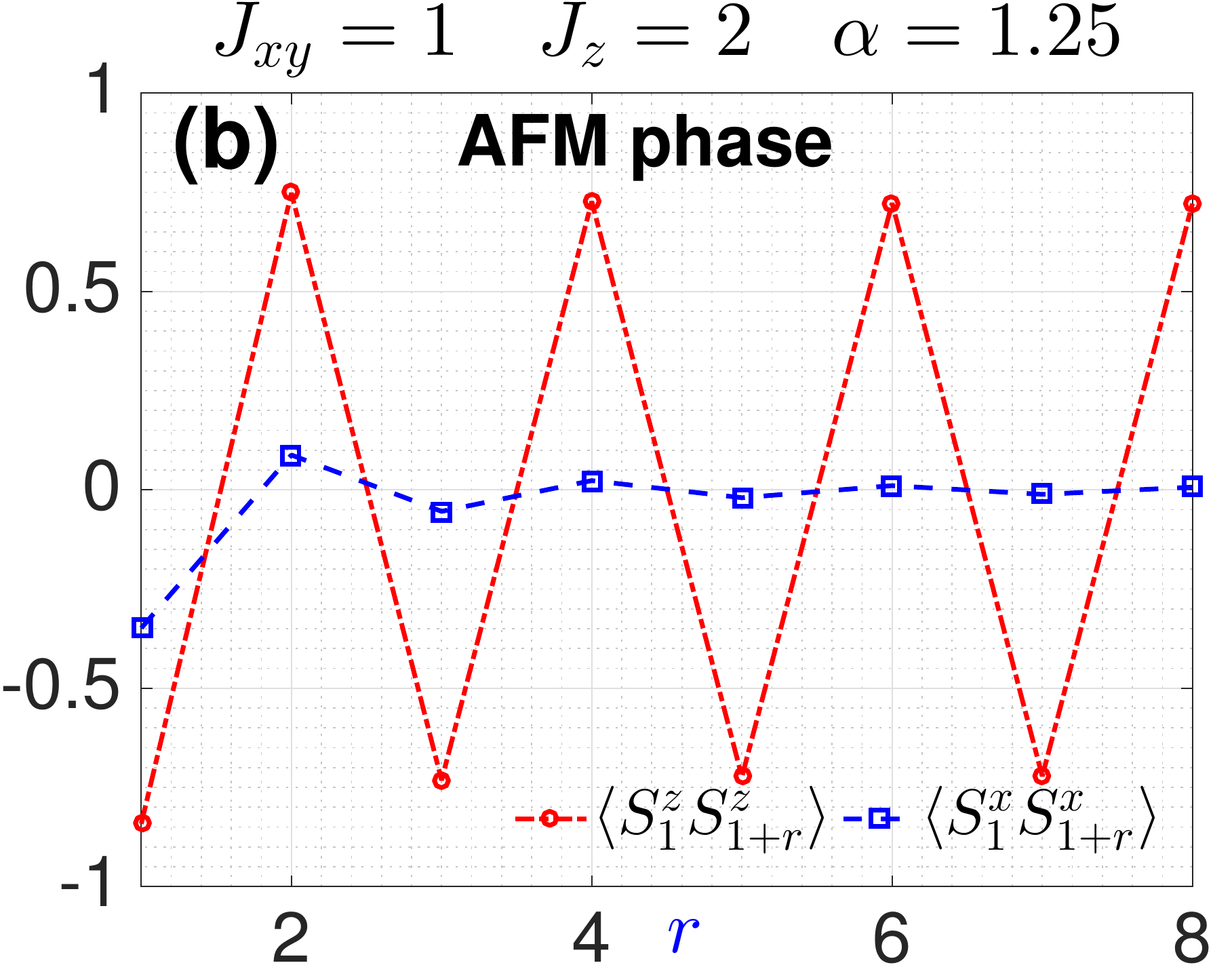}

\medskip{}

\includegraphics[width=0.5\columnwidth]{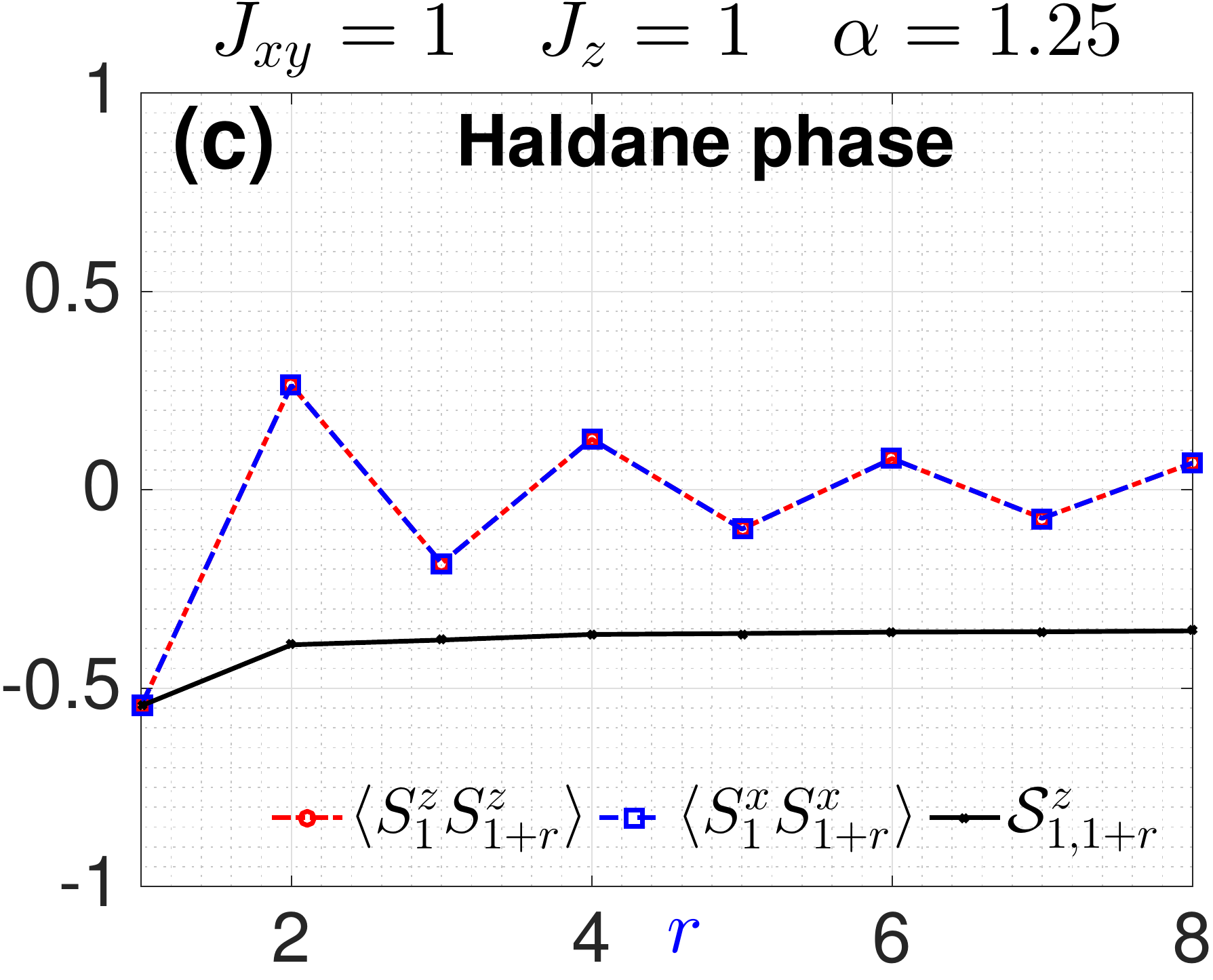}\includegraphics[width=0.5\columnwidth]{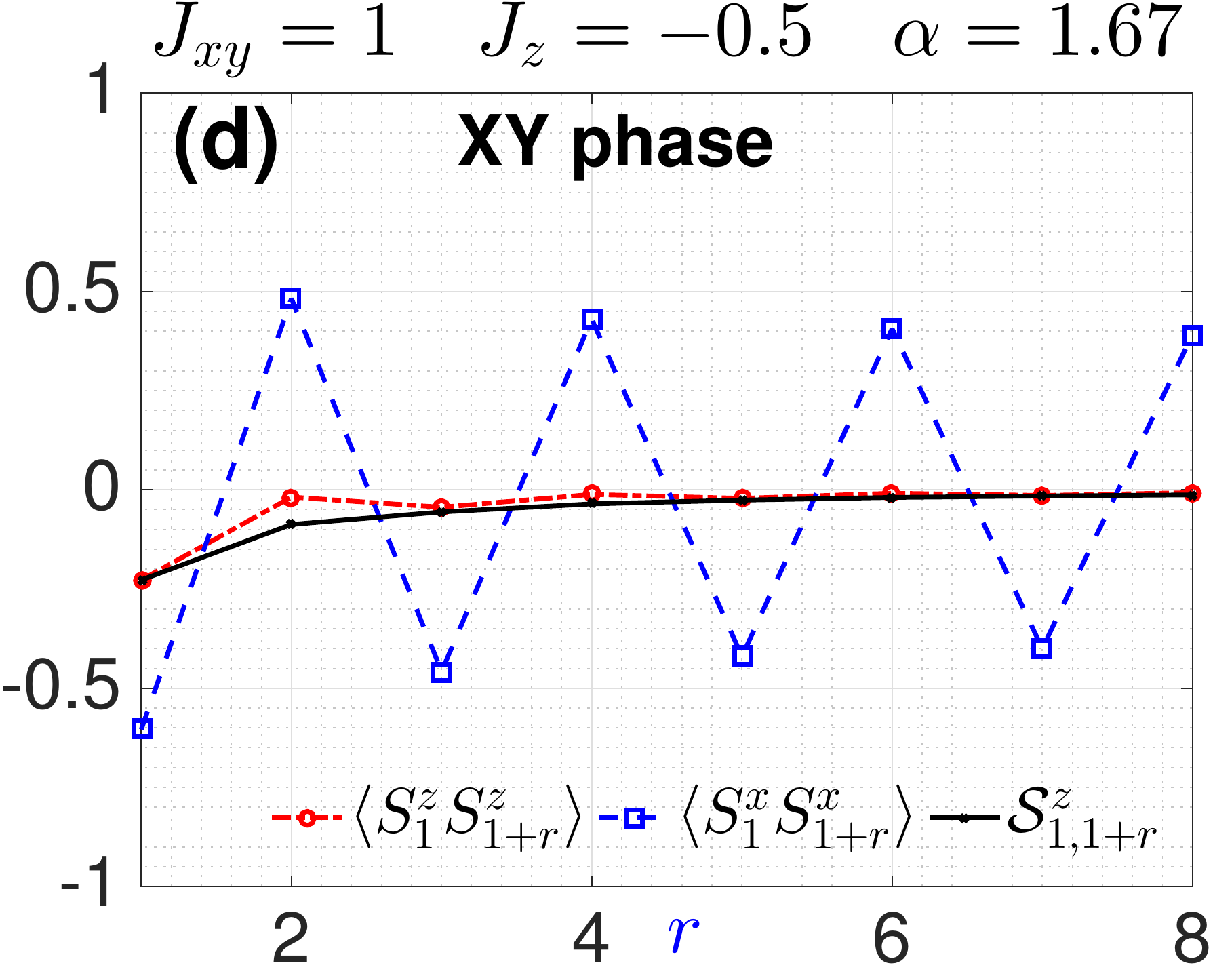}

\medskip{}

\includegraphics[width=0.5\columnwidth]{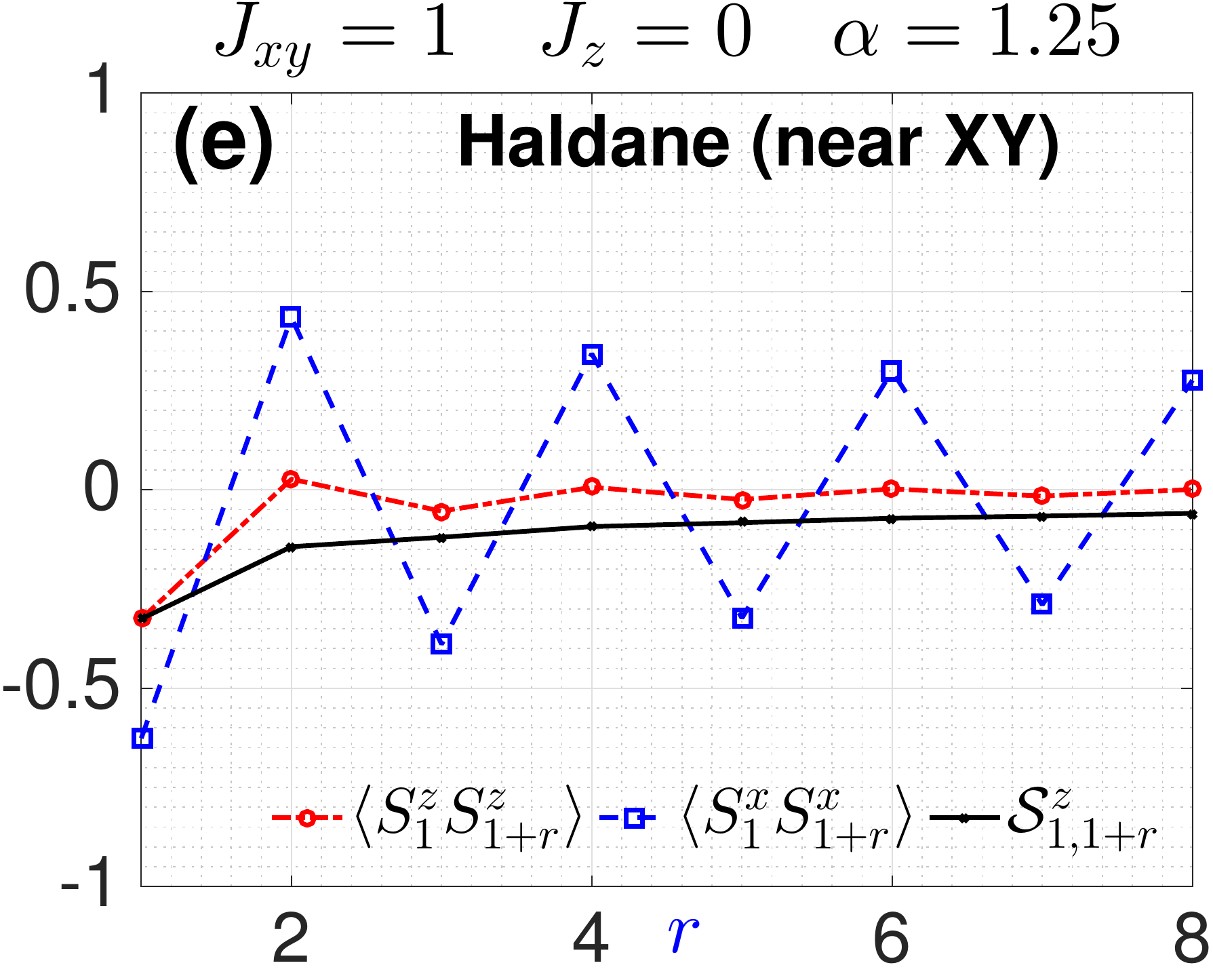}\includegraphics[width=0.5\columnwidth]{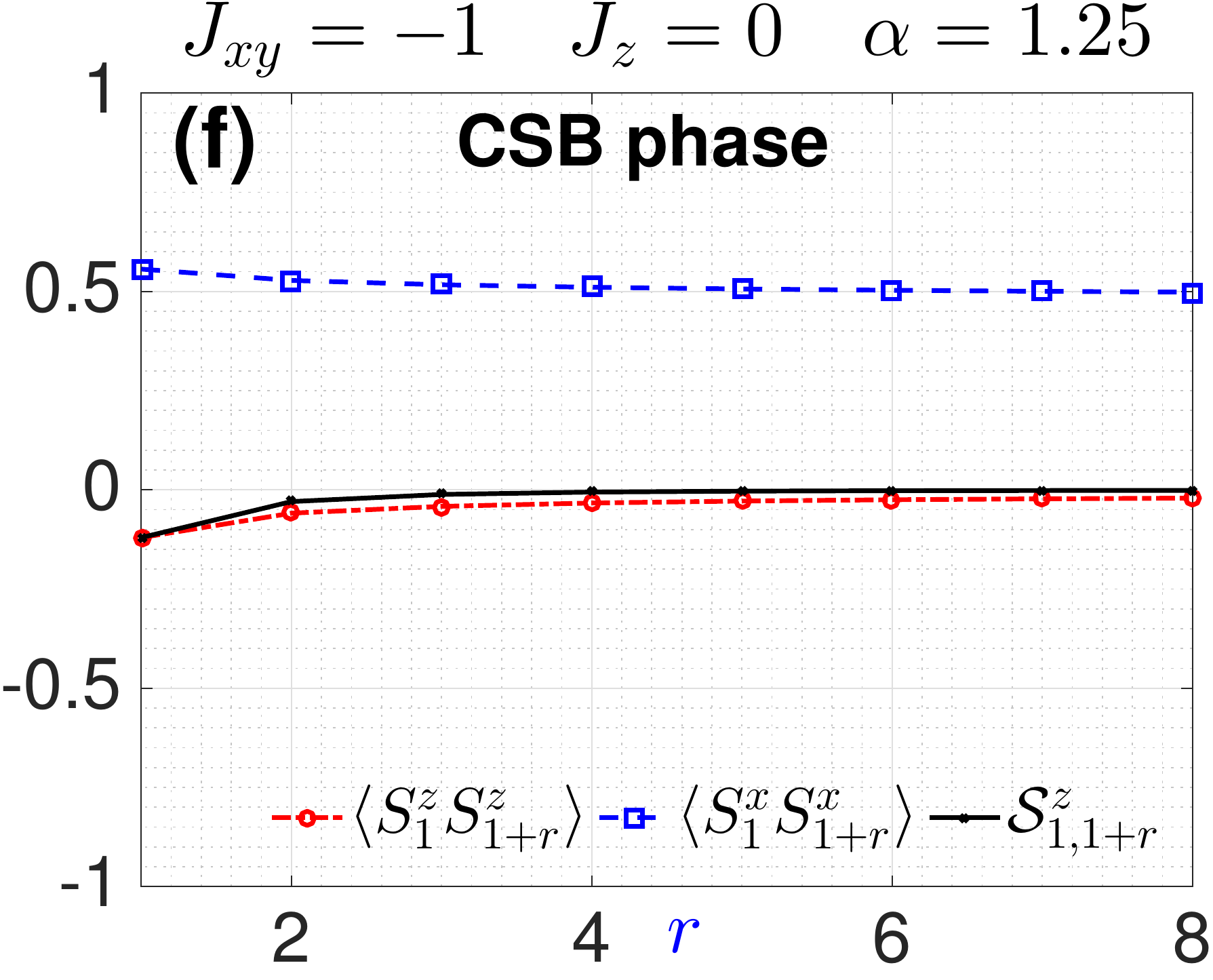}

\caption{Signatures of all five phases for a $N=16$ spin chain. Except for
(e), we tune $J_{xy}$, $J_{z}$ and $\alpha$ to set the ground state
deep into each phase. Each phase is distinguished from the other phases
by different behaviors in various spin-spin correlation functions.\label{fig:Exp}}
\end{figure}

It was theoretically proposed in Refs.\,\cite{cohen_proposal_2014,cohen_simulating_2015}
that the Hamiltonian we consider can be simulated (for widely tunable
$J_{z}$ and $0<\alpha<3$) by using microwave field gradients or
optical dipole forces to induce spin-spin interactions in a chain
of trapped ions. The simulation of Eq.\,\eqref{eq:H} with $J_{xy}=1$
and $J_{z}=0$ was experimentally demonstrated for a few ions with
$\alpha$ tuned around $1$ \cite{senko_realization_2015}, where
the ground state was adiabatically prepared by slowly ramping down
an extra single-ion anisotropy term $D(t)\sum_{i}(S_{i}^{z})^{2}$,
with $D(t)>0$. As the system size increases, the energy gap separating
the ground state from the rest of the spectrum will become progressively
smaller near the point where a phase transition between the ``large-$D$''
phase and the XY/Haldane/FM/AFM phase occurs in the thermodynamic
limit \cite{chen_groundstate_2003}. To avoid a slow ground state
preparation process, we can adiabatically ramp down a staggered magnetic
field in the $z$ direction, $h(t)\sum_{i=1}^{N}(-1)^{i}S_{i}^{z}$,
with $h(t)>0$ \cite{cohen_proposal_2014,cohen_simulating_2015}.
By preparing an initial state that is the highest excited state of
the staggered field Hamiltonian, the same adiabatically ramping process
will lead us to the ground state of the Hamiltonian Eq.\,(\ref{eq:H})
with the opposite sign of both $J_{xy}$ and $J_{z}$. As discussed
in Ref.\,\cite{cohen_simulating_2015}, the spin correlation functions
$\langle S_{i}^{z}S_{j}^{z}\rangle$ and the string-order correlation
$\mathcal{S}_{ij}^{z}\equiv\langle S_{i}^{z}S_{j}^{z}\prod_{i<k<j}(-1)^{S_{k}^{z}}\rangle$
can be measured in trapped-ion experiments for any two ions $i$ and
$j$. Together with arbitrary single-spin rotations performed with
microwave or optical Raman transitions, we can measure these correlations
along any direction. Near-future experiments will most likely be limited
to a few tens of spins. Although this limitation makes it difficult
to probe continuous phase transitions, one can nevertheless observe
important signatures of all five phases discussed in the manuscript
by tuning $J_{z}/J_{xy}$ and $\alpha$ deep into each phase. These
signatures are summarized below and shown in Fig.\,\ref{fig:Exp}.

\
 \textbf{FM phase} {[}Fig.\,\ref{fig:Exp}(a){]}: Within the FM phase,
$\langle S_{i}^{z}S_{j}^{z}\rangle=1$ and $\langle S_{i}^{x}S_{j}^{x}\rangle=0$
for any $i$ and $j$, thus confirming perfect alignment of spins
along the $z$ direction.

\
 \textbf{AFM phase} {[}Fig.\,\ref{fig:Exp}(b){]}: For sufficiently
large $J_{z}>0$, we have $\langle S_{i}^{z}S_{j}^{z}\rangle\approx(-1)^{i-j}$,
showing a near perfect anti-alignment of spins along the $z$ direction.
In contrast, $\langle S_{i}^{x}S_{j}^{x}\rangle$ vanishes over a
separation of just a few sites.

\
 \textbf{Haldane phase} {[}Fig.\,\ref{fig:Exp}(c){]}: $\mathcal{S}_{ij}^{z}$
converges quickly to a nonzero constant as $|i-j|$ increases. In
contrast, $\langle S_{i}^{z}S_{j}^{z}\rangle$ and $\langle S_{i}^{x}S_{j}^{x}\rangle$
vanishe over a separation of just a few sites.

\
 \textbf{XY phase} {[}Fig.\,\ref{fig:Exp}(d){]}: We consider the
XY phase for $J_{xy}=1$ since the $XY$ phase hardly exist for $\alpha<3$
and $J_{xy}=-1$. $\mathcal{S}_{ij}^{z}$ and $\langle S_{i}^{z}S_{j}^{z}\rangle$
both decay quickly to zero as $|i-j|$ increases. $\langle S_{i}^{x}S_{j}^{x}\rangle$
oscillates and its amplitude decays very slowly (the slow decay reflects
a relatively small value of the critical exponent associated with
the correlation function decay).

\
 \textbf{CSB phase} {[}Fig.\,\ref{fig:Exp}(f){]}: As in the XY phase,
both $\mathcal{S}_{ij}^{z}$ and $\langle S_{i}^{z}S_{j}^{z}\rangle$
decay quickly to zero. However, $\langle S_{i}^{x}S_{j}^{x}\rangle$
converges quickly to approximately $0.5$ at large $|i-j|$, showing
a near perfect ordering of spins in the $x-y$ plane. Note that we
are not explicitly breaking $U(1)$ symmetry here, so $\langle S_{i}^{x}S_{j}^{x}\rangle=\langle S_{i}^{y}S_{j}^{y}\rangle=\frac{1}{2}\langle S_{i}^{+}S_{j}^{-}\rangle$.
This is done because it is desirable for the experiment to operate
within the $\sum_{i=1}^{N}S_{i}^{z}=0$ subspace, where magnetic field
noise and unwanted phonon couplings are suppressed \cite{cohen_simulating_2015,senko_realization_2015}.

\
 Finally, we point out that, even in the experimental setup already
demonstrated in Ref.\ \cite{senko_realization_2015}, for which $J_{z}=0$,
one can still explore the two most interesting phases studied in this
manuscript: the Haldane phase and the CSB phase. Note that, for $J_{xy}=1$,
$J_{z}=0$ lies close to the Haldane-to-XY phase boundary, and thus
one observes signatures of both phases, as in Fig.\,\ref{fig:Exp}(e).

\section{Conclusion and Outlook}

By tuning the anisotropy $J_{z}/|J_{xy}|$ and the power-law exponent
$\alpha$, we have explored a rich variety of quantum phases\textemdash and
the transitions between them\textemdash in a long-range interacting
spin-1 XXZ chain. For $J_{xy}=-1$, long-range interactions give rise
to a rather unusual phase diagram due to the emergence of a continuous
symmetry breaking phase in one spatial dimension. Because the CSB
phase cannot happen in short-range interacting 1D spin-system, the
nature of the phase transitions into and out of it is rather interesting;
an in-depth study of the universality class of the CSB-to-XY transition
was carried out in a separate work \cite{maghrebi_continuous_2015},
where a similar transition in the long-range interacting spin-1/2
XXZ chain is analyzed. On the other hand, the CSB-to-Haldane transition,
absent in spin-1/2 chains, requires further study to be understood
thoroughly. The CSB-Haldane-AFM tricritical point is reminiscent of
the tricritical point at the intersection of the large-$D$, Haldane
and AFM phases, which has been related to the integrable Takhtajan-Babujian
model described by an $SU(2)_{2}$ Wess-Zumino-Witten (WZW) model
with central charge $c=3/2$ \cite{tsvelik_fieldtheory_1990,kitazawa_bifurcation_1999,degliespostiboschi_c1_2003,Pixley_frustration_2014,ejima_comparative_2015}.
Additional numerical calculations are needed to accurately determine
the central charge at the CSB-Haldane-AFM tricritical point. Generalizations
of our model to include single-ion anisotropy and a magnetic field
are readily achievable in current trapped-ion experiments \cite{senko_realization_2015,cohen_simulating_2015}.
Understanding these exotic quantum phase transitions\textemdash induced
by long-range interactions that are highly tunable in current experiments\textemdash requires
the confrontation of numerous theoretical and numerical challenges,
and motivates experimental quantum simulation of the model using AMO
systems.

\section*{Acknowledgements}

We thank G.\ Pupillo, D.\,Vodola, L.\, Lepori, A.\,Turner, J.\,Pixley,
M.\ Wall, P.\ Hess, A.\ Lee, J.\ Smith, A.\,Retzker and I.\,Cohen
for helpful discussions. This work was supported by the ARO, the AFOSR,
NSF PIF, NSF PFC at the JQI, and the ARL. M. F.-F. thanks the NRC
for support. 

\bibliographystyle{apsrev4-1}
\bibliography{sub}

\begin{thebibliography}{92}
\bibitem [{Inf()}]{InfKac}%
  \BibitemOpen
  \href@noop {} {\bibinfo  {journal} {For an infinite system, the energy
  renormalization may change the system from gapped to gapless\kern-0.25em}\
  }\BibitemShut {NoStop}%
\bibitem [{Non()}]{NonIntegerAlpha}%
  \BibitemOpen
  \href@noop {} {\bibinfo  {journal} {Here the expression are for non-integer
  $\alpha$. For integer $\alpha$, the expressions are slightly different but
  cause no important changes to the discussion below\kern-0.25em}\
  }\BibitemShut {NoStop}%
\end{thebibliography}

\end{document}